\pdfoutput=1

\documentclass[12pt,a4paper]{article}

\usepackage{ifthen} 
\newboolean{pdflatex}
\setboolean{pdflatex}{true} 

\newboolean{articletitles}
\setboolean{articletitles}{true} 

\newboolean{uprightparticles}
\setboolean{uprightparticles}{false} 

\newboolean{inbibliography}
\setboolean{inbibliography}{false} 

\usepackage{microtype}
\usepackage{lineno}  
\usepackage{xspace} 
\usepackage{caption} 

\usepackage{graphicx}  
\usepackage{color}
\usepackage{colortbl}
\graphicspath{{./figs/}} 

\usepackage{amsmath} 
\usepackage{amssymb}
\usepackage{amsfonts}
\usepackage{upgreek} 

\newcommand*\patchAmsMathEnvironmentForLineno[1]{%
\expandafter\let\csname old#1\expandafter\endcsname\csname #1\endcsname
\expandafter\let\csname oldend#1\expandafter\endcsname\csname
end#1\endcsname
 \renewenvironment{#1}%
   {\linenomath\csname old#1\endcsname}%
   {\csname oldend#1\endcsname\endlinenomath}%
}
\newcommand*\patchBothAmsMathEnvironmentsForLineno[1]{%
  \patchAmsMathEnvironmentForLineno{#1}%
  \patchAmsMathEnvironmentForLineno{#1*}%
}
\AtBeginDocument{%
\patchBothAmsMathEnvironmentsForLineno{equation}%
\patchBothAmsMathEnvironmentsForLineno{align}%
\patchBothAmsMathEnvironmentsForLineno{flalign}%
\patchBothAmsMathEnvironmentsForLineno{alignat}%
\patchBothAmsMathEnvironmentsForLineno{gather}%
\patchBothAmsMathEnvironmentsForLineno{multline}%
\patchBothAmsMathEnvironmentsForLineno{eqnarray}%
}

\usepackage{hyperref}    
\usepackage[all]{hypcap} 

\def\lhcb {\mbox{LHCb}\xspace}

\def\MagUp {\mbox{\em Mag\kern -0.05em Up}\xspace}

\ifthenelse{\boolean{uprightparticles}}%
{

 \def\Pmu         {\ensuremath{\upmu}\xspace}

 \def\Ppi         {\ensuremath{\uppi}\xspace}

 \def\Ppsi        {\ensuremath{\uppsi}\xspace}

 \def\PDelta      {\ensuremath{\Delta}\xspace}                 
 \def\PXi      {\ensuremath{\Xi}\xspace}                 
 \def\PLambda      {\ensuremath{\Lambda}\xspace}                 
 \def\PSigma      {\ensuremath{\Sigma}\xspace}                 
 \def\POmega      {\ensuremath{\Omega}\xspace}                 
 \def\PUpsilon      {\ensuremath{\Upsilon}\xspace}                 
 
 \def\PB      {\ensuremath{\mathrm{B}}\xspace}                 
                  
 \def\PD      {\ensuremath{\mathrm{D}}\xspace}

 \def\PJ      {\ensuremath{\mathrm{J}}\xspace}                 
 \def\PK      {\ensuremath{\mathrm{K}}\xspace}

 \def\PW      {\ensuremath{\mathrm{W}}\xspace}

 \def\Pb      {\ensuremath{\mathrm{b}}\xspace}                 
 \def\Pc      {\ensuremath{\mathrm{c}}\xspace}

 \def\Pi      {\ensuremath{\mathrm{i}}\xspace}

 \def\Pp      {\ensuremath{\mathrm{p}}\xspace}

 \def\Ps      {\ensuremath{\mathrm{s}}\xspace}

}
{

 \def\Pmu         {\ensuremath{\mu}\xspace}

 \def\Ppi         {\ensuremath{\pi}\xspace}

 \def\Ppsi        {\ensuremath{\psi}\xspace}                 
                  
 \mathchardef\PDelta="7101
 \mathchardef\PXi="7104
 \mathchardef\PLambda="7103
 \mathchardef\PSigma="7106
 \mathchardef\POmega="710A
 \mathchardef\PUpsilon="7107
                  
 \def\PB      {\ensuremath{B}\xspace}                 
                  
 \def\PD      {\ensuremath{D}\xspace}

 \def\PJ      {\ensuremath{J}\xspace}                 
 \def\PK      {\ensuremath{K}\xspace}

 \def\PW      {\ensuremath{W}\xspace}

 \def\Pb      {\ensuremath{b}\xspace}                 
 \def\Pc      {\ensuremath{c}\xspace}

 \def\Pi      {\ensuremath{i}\xspace}

 \def\Pp      {\ensuremath{p}\xspace}

 \def\Ps      {\ensuremath{s}\xspace}

}

\makeatletter
\ifcase \@ptsize \relax
  \newcommand{\miniscule}{\@setfontsize\miniscule{4}{5}}
\or
  \newcommand{\miniscule}{\@setfontsize\miniscule{5}{6}}
\or
  \newcommand{\miniscule}{\@setfontsize\miniscule{5}{6}}
\fi
\makeatother

\DeclareRobustCommand{\optbar}[1]{\shortstack{{\miniscule (\rule[.5ex]{1.25em}{.18mm})}
  \\ [-.7ex] $#1$}}

\def\mup        {{\ensuremath{\Pmu^+}}\xspace}
\def\mun        {{\ensuremath{\Pmu^-}}\xspace} 
\def\mumu       {{\ensuremath{\Pmu^+\Pmu^-}}\xspace}

\def\Wpm    {{\ensuremath{\PW^\pm}}\xspace}

\def\squark    {{\ensuremath{\Ps}}\xspace}

\def\cquark    {{\ensuremath{\Pc}}\xspace}
\def\cquarkbar {{\ensuremath{\overline \cquark}}\xspace}

\def\bquark    {{\ensuremath{\Pb}}\xspace}

\def\pion   {{\ensuremath{\Ppi}}\xspace}

\def\pip    {{\ensuremath{\pion^+}}\xspace}
\def\pim    {{\ensuremath{\pion^-}}\xspace}

\def\kaon    {{\ensuremath{\PK}}\xspace}
  \def\Kbar    {{\kern 0.2em\overline{\kern -0.2em \PK}{}}\xspace}

\def\KorKbar    {\kern 0.18em\optbar{\kern -0.18em K}{}\xspace}

\def\KS      {{\ensuremath{\kaon^0_{\rm\scriptscriptstyle S}}}\xspace}

  \def\Dbar    {{\kern 0.2em\overline{\kern -0.2em \PD}{}}\xspace}

\def\DorDbar    {\kern 0.18em\optbar{\kern -0.18em D}{}\xspace}

\def\B       {{\ensuremath{\PB}}\xspace}
\def\Bbar    {{\ensuremath{\kern 0.18em\overline{\kern -0.18em \PB}{}}}\xspace}

\def\BorBbar    {\kern 0.18em\optbar{\kern -0.18em B}{}\xspace}
\def\Bz      {{\ensuremath{\B^0}}\xspace}

\def\jpsi     {{\ensuremath{{\PJ\mskip -3mu/\mskip -2mu\Ppsi\mskip 2mu}}}\xspace}
\def\psitwos  {{\ensuremath{\Ppsi{(2S)}}}\xspace}

  \def\Y#1S{\ensuremath{\PUpsilon{(#1S)}}\xspace}

\def\proton      {{\ensuremath{\Pp}}\xspace}

\def\Lz          {{\ensuremath{\PLambda}}\xspace}
\def\Lbar        {{\ensuremath{\kern 0.1em\overline{\kern -0.1em\PLambda}}}\xspace}
\def\LorLbar    {\kern 0.18em\optbar{\kern -0.18em \PLambda}{}\xspace}

\def\Lb      {{\ensuremath{\Lz^0_\bquark}}\xspace}
\def\Lbbar   {{\ensuremath{\Lbar{}^0_\bquark}}\xspace}

\def\BF         {{\ensuremath{\cal B}}\xspace}

\newcommand{\decay}[2]{\ensuremath{#1\!\to #2}\xspace}         

\def\to                 {\ensuremath{\rightarrow}\xspace}

\def\qsq       {{\ensuremath{q^2}}\xspace}

\def\CP                {{\ensuremath{C\!P}}\xspace}

\newcommand{\etot}{{\ensuremath{\varepsilon_{\rm tot}}}\xspace}

\def\AT#1     {\ensuremath{A_{\mathrm{T}}^{#1}}\xspace}           

\def\C#1      {\ensuremath{\mathcal{C}_{#1}}\xspace}                       
\def\Cp#1     {\ensuremath{\mathcal{C}_{#1}^{'}}\xspace}                    
\def\Ceff#1   {\ensuremath{\mathcal{C}_{#1}^{\mathrm{(eff)}}}\xspace}        
\def\Cpeff#1  {\ensuremath{\mathcal{C}_{#1}^{'\mathrm{(eff)}}}\xspace}       
\def\Ope#1    {\ensuremath{\mathcal{O}_{#1}}\xspace}                       
\def\Opep#1   {\ensuremath{\mathcal{O}_{#1}^{'}}\xspace}                    

\newcommand{\tev}{\ifthenelse{\boolean{inbibliography}}{\ensuremath{~T\kern -0.05em eV}\xspace}{\ensuremath{\mathrm{\,Te\kern -0.1em V}}}\xspace}
\newcommand{\gev}{\ensuremath{\mathrm{\,Ge\kern -0.1em V}}\xspace}
\newcommand{\mev}{\ensuremath{\mathrm{\,Me\kern -0.1em V}}\xspace}
\newcommand{\kev}{\ensuremath{\mathrm{\,ke\kern -0.1em V}}\xspace}
\newcommand{\ev}{\ensuremath{\mathrm{\,e\kern -0.1em V}}\xspace}
\newcommand{\gevc}{\ensuremath{{\mathrm{\,Ge\kern -0.1em V\!/}c}}\xspace}
\newcommand{\mevc}{\ensuremath{{\mathrm{\,Me\kern -0.1em V\!/}c}}\xspace}
\newcommand{\gevcc}{\ensuremath{{\mathrm{\,Ge\kern -0.1em V\!/}c^2}}\xspace}
\newcommand{\gevgevcccc}{\ensuremath{{\mathrm{\,Ge\kern -0.1em V^2\!/}c^4}}\xspace}
\newcommand{\mevcc}{\ensuremath{{\mathrm{\,Me\kern -0.1em V\!/}c^2}}\xspace}

\def\mum  {\ensuremath{{\,\upmu\rm m}}\xspace}

\def\invfb   {\ensuremath{\mbox{\,fb}^{-1}}\xspace}

\def\ps   {\ensuremath{{\rm \,ps}}\xspace}

\newcommand{\chisq}{\ensuremath{\chi^2}\xspace}

\def\deriv {\ensuremath{\mathrm{d}}}

\def\gsim{{~\raise.15em\hbox{$>$}\kern-.85em
          \lower.35em\hbox{$\sim$}~}\xspace}
\def\lsim{{~\raise.15em\hbox{$<$}\kern-.85em
          \lower.35em\hbox{$\sim$}~}\xspace}

\def\ptot       {\mbox{$p$}\xspace}
\def\pt         {\mbox{$p_{\rm T}$}\xspace}

\def\mrad{\ensuremath{\rm \,mrad}\xspace}

\def\evtgen     {\mbox{\textsc{EvtGen}}\xspace}

\def\geant      {\mbox{\textsc{Geant4}}\xspace}

\def\photos     {\mbox{\textsc{Photos}}\xspace}

\def\pythia     {\mbox{\textsc{Pythia}}\xspace}

\def\tell1  {TELL1\xspace}
\def\ukl1   {UKL1\xspace}

\def\AFBl {\ensuremath{A^{\ell}_{\mathrm{FB}}}\xspace}
\def\AFBh {\ensuremath{A^{h}_{\mathrm{FB}}}\xspace}

\def\thetal {\ensuremath{\theta_\ell}\xspace}

\usepackage{cite} 
\usepackage{mciteplus}

\usepackage{longtable} 

\begin{document}

\renewcommand{\thefootnote}{\fnsymbol{footnote}}
\setcounter{footnote}{1}


\begin{titlepage}
\pagenumbering{roman}

\vspace*{-1.5cm}
\centerline{\large EUROPEAN ORGANIZATION FOR NUCLEAR RESEARCH (CERN)}
\vspace*{1.5cm}
\hspace*{-0.5cm}
\begin{tabular*}{\linewidth}{lc@{\extracolsep{\fill}}r}
\ifthenelse{\boolean{pdflatex}}
{\vspace*{-2.7cm}\mbox{\!\!\!\includegraphics[width=.14\textwidth]{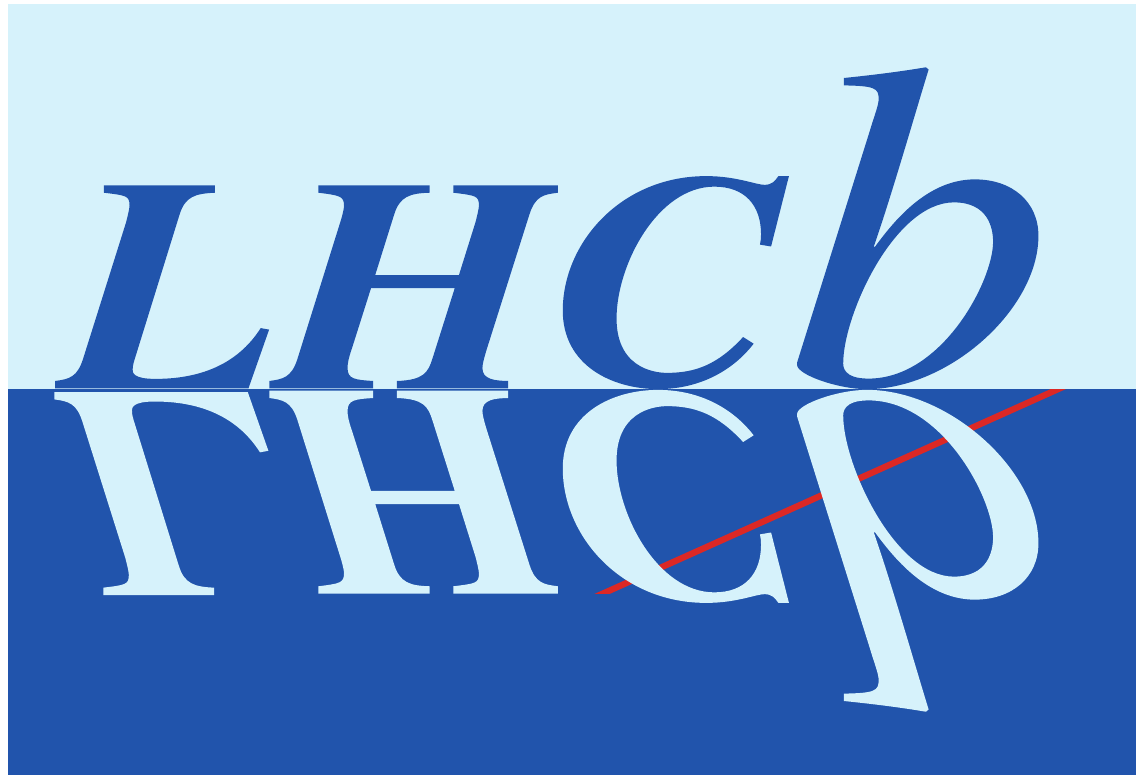}} & &}%
{\vspace*{-1.2cm}\mbox{\!\!\!\includegraphics[width=.12\textwidth]{lhcb-logo.eps}} & &}%
\\
 & & CERN-PH-EP-2015-078 \\  
 & & LHCb-PAPER-2015-009 \\  
 & & 24 March 2015  \\
\end{tabular*}

\vspace*{1.5cm}

{\bf\boldmath\huge
\begin{center}
  Differential branching fraction
  and angular analysis of
  \decay{\Lb}{\Lz\mumu} decays
\end{center}
}

\vspace*{0.7cm}

\begin{center}
The LHCb collaboration\footnote{Authors are listed at the end of this paper.}
\end{center}


\begin{abstract}
  \noindent
  The differential branching fraction of the rare decay
  \decay{\Lb}{\Lz\mumu} is measured as a function of \qsq, the square
  of the dimuon invariant mass.  The analysis is performed using
  proton-proton collision data, corresponding to an integrated
  luminosity of 3.0\invfb, collected by the \lhcb experiment. 
  Evidence of signal is observed in the \qsq region below the square
  of the \jpsi mass. Integrating
  over $15 < \qsq < 20$ \gevgevcccc 
 the branching fraction
 is measured as
\begin{equation}
\deriv\BF(\decay{\Lb}{\Lz\mumu})/\deriv\qsq = (1.18 \;^{+\,0.09}_{-\,0.08} \pm 0.03 \pm 0.27 )
\times 10^{-7}\;(\gevgevcccc)^{-1},
\nonumber
\end{equation}
 \noindent where the uncertainties are statistical, 
 systematic and due to the normalisation mode,
 \decay{\Lb}{\jpsi\Lz}, respectively.
 In the \qsq\ intervals where the signal is observed, angular
 distributions are studied and the forward-backward asymmetries
 in the dimuon ($A_{\rm FB}^{\ell}$) and hadron ($A_{\rm FB}^{h})$ systems
 are measured for the first time.
 In the range $15 < \qsq < 20$ \gevgevcccc they are found to be\footnote{Please see erratum in appendix \ref{sec:erratum}}
\begin{equation}
\begin{split}
A_{\rm FB}^{\ell} & = -0.05 \; \pm 0.09 \; \text{(stat)} \; \pm  0.03 \; \text{(syst)} \text{\;and} \\
A_{\rm FB}^{h} & = -0.29 \; \pm 0.07 \; \text{(stat)} \; \pm  0.03 \; \text{(syst)}.
\end{split}
\nonumber
\end{equation}

\end{abstract}

\vspace*{1.0cm}

\begin{center}
  Published in JHEP 06 (2015) 115, JHEP 09 (2018) 145.
\end{center}

\vspace{\fill}

{\footnotesize 
\centerline{\copyright~CERN on behalf of the \lhcb collaboration, licence \href{http://creativecommons.org/licenses/by/4.0/}{CC-BY-4.0}.}}
\vspace*{2mm}

\end{titlepage}


\newpage
\setcounter{page}{2}
\mbox{~}

\cleardoublepage


\renewcommand{\thefootnote}{\arabic{footnote}}
\setcounter{footnote}{0}



\pagestyle{plain} 
\setcounter{page}{1}
\pagenumbering{arabic}


%

\section{Introduction}
The decay \decay{\Lb}{\Lz\mumu} is a rare (\decay{\bquark}{\squark})
flavour-changing neutral current process that, in the Standard Model
(SM), proceeds through electroweak loop (penguin and \Wpm box)
diagrams.  As non-SM particles may also contribute to the decay
amplitudes, measurements of this and similar decays can be used to
search for physics beyond the SM. To date, emphasis has been placed on
the study of rare decays of mesons rather than baryons, in part due to
the theoretical complexity of the latter \cite{Mannel:1997xy}.  In the
particular system studied in this analysis, the decay products include
only a single long-lived hadron, simplifying the theoretical modelling
of hadronic physics in the final state.

The study of \Lb baryon decays is of considerable interest for several
reasons.  Firstly, as the \Lb baryon {has non-zero spin}, there is the
potential to improve the limited understanding of the helicity
structure of the underlying Hamiltonian, which cannot be extracted
from meson decays \cite{Hiller:2007ur,Mannel:1997xy}. Secondly, as the
\Lb baryon may be considered as consisting of a heavy quark combined
with a light diquark system, the hadronic physics differs
significantly from that of the \B meson decay.  A further motivation
specific to the \decay{\Lb}{\Lz\mumu} channel is that the polarisation
of the \Lz baryon is preserved in the \decay{\Lz}{\proton\pim}
decay\footnote{The inclusion of charge-conjugate modes is implicit
  throughout.}, giving access to complementary information to that
available from meson decays \cite{Boer:2014kda}.

Theoretical aspects of the \decay{\Lb}{\Lz\mumu} decay have been
considered both in the SM and in some of its extensions
\cite{Boer:2014kda,Aslam:2008hp,Wang:2008sm,Huang:1998ek,Chen:2001ki,Chen:2001zc,
  Chen:2001sj,Zolfagharpour:2007eh,Mott:2011cx,Aliev:2010uy,Mohanta:2010eb,Sahoo:2011yb,Detmold:2012vy,Gutsche:2013pp}.
Although based on the same effective Hamiltonian as that for the
corresponding mesonic transitions, the hadronic form factors for the
\Lb baryon case are less well-known due to the less stringent
experimental constraints.  This leads to a large spread in the
predicted branching fractions.  The decay has a non-trivial angular
structure which, in the case of unpolarised \Lb production, is
described by the helicity angles of the muon and proton, the angle
between the planes defined by the \Lz decay products and the two
muons, and the square of the dimuon invariant mass, \qsq.  In
theoretical investigations, the differential branching fraction, and
forward-backward asymmetries for both the dilepton and the hadron
systems of the decay, have received particular attention
\cite{Boer:2014kda,Mott:2011cx,Detmold:2012vy,Meinel:2014wua,Gutsche:2013pp}.
Different treatments of form factors are used depending on the \qsq
region and can be tested by comparing predictions with data as a
function of \qsq.
 
In previous observations of the decay \decay{\Lb}{\Lz\mumu}
\cite{Aaltonen:2011qs,LHCB-PAPER-2013-025}, evidence for signal had
been limited to \qsq values above the square of the mass of the
\psitwos resonance. This region will be referred to as ``high-\qsq'',
while that below the \psitwos will be referred to as ``low-\qsq''.  In
this paper an updated measurement by \lhcb of the differential
branching fraction for the rare decay \decay{\Lb}{\Lz\mumu}, and the
first angular analysis of this decay mode, are reported.
Non-overlapping \qsq intervals in the range 0.1--20.0\gevgevcccc, and
theoretically motivated ranges 1.1--6.0 and 15.0--20.0\gevgevcccc ~\cite{Boer:2014kda,Beylich:2011aq,Beneke:2000wa}, are
used.  The rates are normalised with respect to the tree-level
\decay{\bquark}{\cquark\cquarkbar\squark} decay \decay{\Lb}{\jpsi\Lz},
where \decay{\jpsi}{\mumu}. This analysis uses $pp$ collision data,
corresponding to an integrated luminosity of 3.0\invfb, collected
during 2011 and 2012 at centre-of-mass energies of 7 and 8\tev,
respectively.

\section{Detector and simulation}
\label{sec:Detector}
The \lhcb detector~\cite{Alves:2008zz,LHCb-DP-2014-002} is a
single-arm forward spectrometer covering the \mbox{pseudorapidity}
range $2<\eta <5$, designed for the study of particles containing
\bquark or \cquark quarks. The detector includes a high-precision
tracking system (VELO) consisting of a silicon-strip vertex detector
surrounding the $pp$ interaction region~\cite{LHCb-DP-2014-001}, a
large-area silicon-strip detector located upstream of a dipole magnet
with a bending power of about $4{\rm\,Tm}$, and three stations of
silicon-strip detectors and straw drift tubes~\cite{LHCb-DP-2013-003}
placed downstream of the magnet.  The tracking system provides a
measurement of momentum, \ptot, with a relative uncertainty that
varies from 0.5\% at low momentum to 1.0\% at 200\gevc.  The minimum
distance of a track to a primary vertex, the impact parameter, is
measured with a resolution of $(15+29/\pt)\mum$, where \pt is the
component of the momentum transverse to the beam, in\,\gevc.
Different types of charged hadrons are distinguished using information
from two ring-imaging Cherenkov
(RICH) detectors~\cite{LHCb-DP-2012-003}. Photon, electron and hadron
candidates are identified using a calorimeter system that consists of
scintillating-pad and preshower detectors, an electromagnetic
calorimeter and a hadronic calorimeter. Muons are identified by a
system composed of alternating layers of iron and multiwire
proportional chambers~\cite{LHCb-DP-2012-002}.

The trigger~\cite{LHCb-DP-2012-004} consists of a hardware stage,
based on information from the calorimeter and muon systems, followed
by a software stage in which a full event reconstruction is carried
out.  Candidate events are first required to pass a hardware trigger,
which selects muons with a transverse momentum $\pt>1.48\gevc$ in the
7\tev data or $\pt>1.76\gevc$ in the 8\tev data. In the subsequent
software trigger, at least one of the final-state charged particles is
required to have both $\pt>0.8\gevc$ and impact parameter greater than
$100\mum$ with respect to all of the primary $pp$ interaction
vertices~(PVs) in the event. Finally, the tracks of two or more of the
final-state particles are required to form a vertex that is
significantly displaced from the PVs.

Simulated samples of $pp$ collisions are generated using
\pythia~\cite{Sjostrand:2007gs} with a specific \lhcb
configuration~\cite{LHCb-PROC-2010-056}.  Decays of hadronic particles
are described by \evtgen~\cite{Lange:2001uf}, in which final-state
radiation is generated using \photos~\cite{Golonka:2005pn}. The
interaction of the generated particles with the detector, and its
response, are implemented using the \geant
toolkit~\cite{Allison:2006ve, *Agostinelli:2002hh} as described in
Ref.~\cite{LHCb-PROC-2011-006}.  The model used in the simulation of
\decay{\Lb}{\Lz\mumu} decays includes \qsq and angular dependence as
described in Ref.~\cite{Gutsche:2013pp}, together with Wilson
coefficients based on Refs.~\cite{Buras:1994dj,Buras:1993xp}.
Interference effects from \jpsi and $\psi(2S)$ contributions are not
included. For the \decay{\Lb}{\jpsi\Lz} decay the simulation model is
based on the angular distributions observed in
Ref.~\cite{LHCb-PAPER-2012-057}.

\section{Candidate selection}

 Candidate \decay{\Lb}{\Lz\mumu} (signal mode) and
 \decay{\Lb}{\jpsi\Lz} (normalisation mode) decays are reconstructed
 from a \Lz baryon candidate and either a dimuon or a \jpsi meson
 candidate, respectively.  The \decay{\Lb}{\jpsi\Lz} mode, with the
 \jpsi meson reconstructed via its dimuon decay, is a convenient
 normalisation process because it has the same final-state particles
 as the signal mode. 
 Signal and normalisation channels are distinguished by the
 \qsq interval in which they fall. 

 The dimuon candidates are formed from two
 well-reconstructed oppositely charged particles that are
 significantly displaced from any PV, identified as muons and
 consistent with originating from a common vertex.
 
 Candidate \Lz decays are reconstructed in the
 \decay{\Lz}{\proton\pim} mode from two oppositely charged tracks that
 either both include information from the VELO (\textit{long}
 candidates), or both do not include information from the VELO
 (\textit{downstream} candidates). The \Lz candidates must also have a
 vertex fit with a good \chisq, a decay time of at least 2\ps and an
 invariant mass within 30\mevcc of the known \Lz
 mass~\cite{Agashe:2014kda}.  For long candidates, charged particles
 must have $\pt>0.25\gevc$ and a further requirement is imposed on the
 particle identification (PID) of the proton using a likelihood
 variable that combines information from the RICH detectors and the
 calorimeters.

 Candidate \Lb decays are formed from \Lz and dimuon candidates that
 have a combined invariant mass in the interval 5.3--7.0\gevcc and
 form a good-quality vertex that is well-separated from any PV.
 Candidates pointing to the PV with which they are associated are
 selected by requiring that the angle between the \Lb momentum vector
 and the vector between the PV and the \Lb decay vertex, $\theta_D$,
 is less than 14\mrad.  After the \Lb candidate is built, a kinematic
 fit \cite{Hulsbergen:2005pu} of the complete decay chain is performed
 in which the proton and pion are constrained such that the
 $\proton\pim$ invariant mass corresponds to the known \Lz baryon
 mass, and the \Lz and dimuon systems are constrained to originate
 from their respective vertices. Furthermore, candidates falling in the
 $8-11$ and $12.5-15$ \gevgevcccc ~\qsq intervals are excluded
 from the rare sample as they are dominated by decays via
 \jpsi and \psitwos resonances.

 The final selection is based on a neural network classifier
 \cite{Feindt:2006pm,feindt-2004}, exploiting 15 variables carrying
 kinematic, candidate quality and particle identification information.
 Both the track parameter resolutions and kinematic properties are different
 for downstream and long \Lz decays and therefore a separate training is
 performed for each category.  The signal sample used to train the neural
 network consists of simulated \decay{\Lb}{\Lz\mumu} events, while the
 background is taken from data in the upper sideband of the \Lb
 candidate mass spectrum, between 6.0 and 7.0\gevcc. Candidates with a
 dimuon mass in either the \jpsi or \psitwos regions ($\pm 100$\mevcc
 intervals around their known masses)
 are excluded from the training samples.  The variable that provides the greatest
 discrimination in the case of long candidates is the \chisq from the
 kinematic fit.  For downstream candidates, the \pt of the \Lz
 candidate is the most powerful variable.  Other variables that
 contribute significantly are: the PID information for muons; the
 separation of the muons, the pion and the \Lb\ candidate from the PV;
 the distance between the \Lz and \Lb decay vertices; and the pointing
 angle, $\theta_D$.
 
 The requirement on the response of the neural network classifier is
 chosen separately for low- and high-\qsq candidates using two
 different figures of merit.  In the low-\qsq region, where the signal
 has not been previously established, the figure of merit $\varepsilon
 /(\sqrt{N_{\mathrm{B}}} + a/2)$~\cite{Punzi:2003bu} is used, where
 $\varepsilon$ and $N_{\mathrm{B}}$ are the signal efficiency and the
 expected number of background decays and $a$ is the target
 significance; a value of $a = 3$ is used.  In contrast, for the
 high-\qsq region the figure of merit
 $N_{\mathrm{S}}/\sqrt{N_{\mathrm{S}}+N_{\mathrm{B}}}$ is maximised,
 where $N_{\mathrm{S}}$ is the expected number of signal candidates.
 To ensure an appropriate normalisation of $N_{\mathrm{S}}$, the
 number of \decay{\Lb}{\jpsi\Lz} candidates that satisfy the
 preselection is scaled by the measured ratio of branching fractions
 of \decay{\Lb}{\Lz\mumu} to \decay{\Lb}{\jpsi(\to\mumu)\Lz}
 decays~\cite{LHCB-PAPER-2013-025}, and the \decay{\jpsi}{\mumu}
 branching fraction~\cite{Agashe:2014kda}.  The value of
 $N_\mathrm{B}$ is determined by extrapolating the number of candidate
 decays found in the background training sample into the signal
 region.  Relative to the preselected event sample, the neural network
 retains approximately 96\,\% (66\,\%) of downstream candidates and
 97\,\% (82\,\%) of long candidates for the selection at high (low)
 \qsq.

\section{Peaking backgrounds}
\label{sec:physbg}
 In addition to combinatorial background formed from the random
 combination of particles, backgrounds due to specific decays are
 studied using fully reconstructed samples of simulated \bquark hadron
 decays in which the final state includes two muons.  For the
 \decay{\Lb}{\jpsi\Lz} channel, the only significant contribution is
 from \decay{\Bz}{\jpsi\KS} decays, with \decay{\KS}{\pip\pim} where
 one of the pions is misidentified as a proton.  This decay contains a
 long-lived \KS meson and therefore has the same topology as the
 \decay{\Lb}{\jpsi\Lz} mode. This contribution leads to a broad shape
 that peaks below the \Lb mass region, which is taken into account in
 the mass fit.

 For the \decay{\Lb}{\Lz\mumu} channel two sources of peaking
 background are identified. The first of these is
 \decay{\Lb}{\jpsi\Lz} decays in which an energetic photon is radiated
 from either of the muons; this constitutes a background in the \qsq
 region just below the square of the \jpsi mass and in a mass region
 significantly below the \Lb mass.  These events do not contribute
 significantly in the \qsq intervals chosen for the analysis.  The
 second source of background is due to \decay{\Bz}{\KS\mumu} decays,
 where \decay{\KS}{\pip\pim} and one of the pions is misidentified as
 a proton.  This contribution is estimated by scaling the number of
 \decay{\Bz}{\jpsi\KS} events found in the \decay{\Lb}{\jpsi\Lz} fit
 by the ratio of the world average branching fractions for the decay
 processes \decay{\Bz}{\KS\mumu} and \decay{\Bz}{\jpsi(\to\mumu)\KS}
 \cite{Agashe:2014kda}.  Integrated over \qsq this is estimated to
 yield fewer than ten events, which is small relative to the expected
 total background level.

\section{Yields}

\subsection{Fit procedure}
 The yields of signal and background events in the data are
 determined in the mass range 5.35--6.00\gevcc using unbinned
 extended maximum likelihood fits for the \decay{\Lb}{\Lz\mumu} and
 the \decay{\Lb}{\jpsi\Lz} modes.  The likelihood function has the form
\begin{equation}
\mathcal{L}=e^{-(N_\mathrm{S}+N_\mathrm{C}+N_{\mathrm{P}})} \times \prod_{i=1}^{N}[
  N_\mathrm{S}P_{\mathrm{S}}(m_i)+N_\mathrm{C}P_\mathrm{C}(m_i)+N_{\mathrm{P}}P_{\mathrm{P}}(m_i)]
 \;,
\label{eq:uml}
\end{equation}
 \noindent where $N_\mathrm{S}$, $N_\mathrm{C}$ and $N_\mathrm{P}$ are
 the number of
 signal, combinatorial and peaking background events, respectively, 
 $P_j(m_i)$ are the corresponding probability density functions
 (PDFs) and $m_i$  is the mass of the \Lb candidate. 
 The signal yield itself is parametrised in the fit using the
 relative branching fraction of the signal and normalisation modes,
\begin{equation}
N_\mathrm{S}(\Lz\mumu)_{k}  = \left[ \frac{\mathrm{d}\mathcal{B}(\Lz\mumu)/\mathrm{d}\qsq}{\mathcal{B}(\jpsi\Lz)} \right]  \cdot
N_\mathrm{S}(\jpsi\Lz)_{k} \cdot \varepsilon^{\mathrm{rel}}_{k} \cdot \frac {\Delta\qsq} { \mathcal{B}(\jpsi\to\mumu) },
\label{eq:ield_from_BR}
\end{equation}
\noindent
where $k$ is the candidate category (long or downstream), $\Delta\qsq$
is the width of the \qsq interval considered and
$\varepsilon_k^{\mathrm{rel}}$ is the relative efficiency, fixed to
the values obtained as described in Sec.~\ref{sec:efficiency}. Fitting
the ratio of the branching fractions of signal and normalisation modes
simultaneously in both candidate categories makes better statistical
use of the data.

 The signal shape, in both \decay{\Lb}{\Lz\mumu} and
 \decay{\Lb}{\jpsi\Lz} modes, is described by the sum of two Crystal
 Ball functions~\cite{Skwarnicki:1986xj} that share common means and
 tail parameters but have independent widths.  The combinatorial
 background is parametrised by an exponential function, independently
 in each \qsq interval. The background due to \decay{\Bz}{\jpsi\KS}
 decays is modelled by the sum of two Crystal Ball functions with
 opposite tails. All shape parameters are independent
 for the downstream and long sample.

 For the \decay{\Lb}{\jpsi\Lz} mode, the widths and common mean in the
 signal parametrisation are free parameters. The parameters describing
 the shape of the peaking background are fixed to those derived from
 simulated \decay{\Bz}{\jpsi\KS} decays, with only the normalisation
 allowed to vary to accomodate differences between data and simulation.

 For the \decay{\Lb}{\Lz\mumu} decay, the signal shape parameters are
 fixed according to the result of the fit to \decay{\Lb}{\jpsi\Lz}
 data and the widths are rescaled to allow for possible differences
 in resolution as a function of \qsq. The scaling factor is determined
 comparing \decay{\Lb}{\jpsi\Lz} and \decay{\Lb}{\Lz\mumu} simulated events.
 The \decay{\Bz}{\KS\mumu} background component is also
 modelled using the sum of two Crystal Ball functions with opposite
 tails where both the yield and all shape parameters are constrained
 to those obtained from simulated events.
 
\subsection{Fit results}
\begin{figure}[tpbh!]
\centering \includegraphics[width=0.8\textwidth]{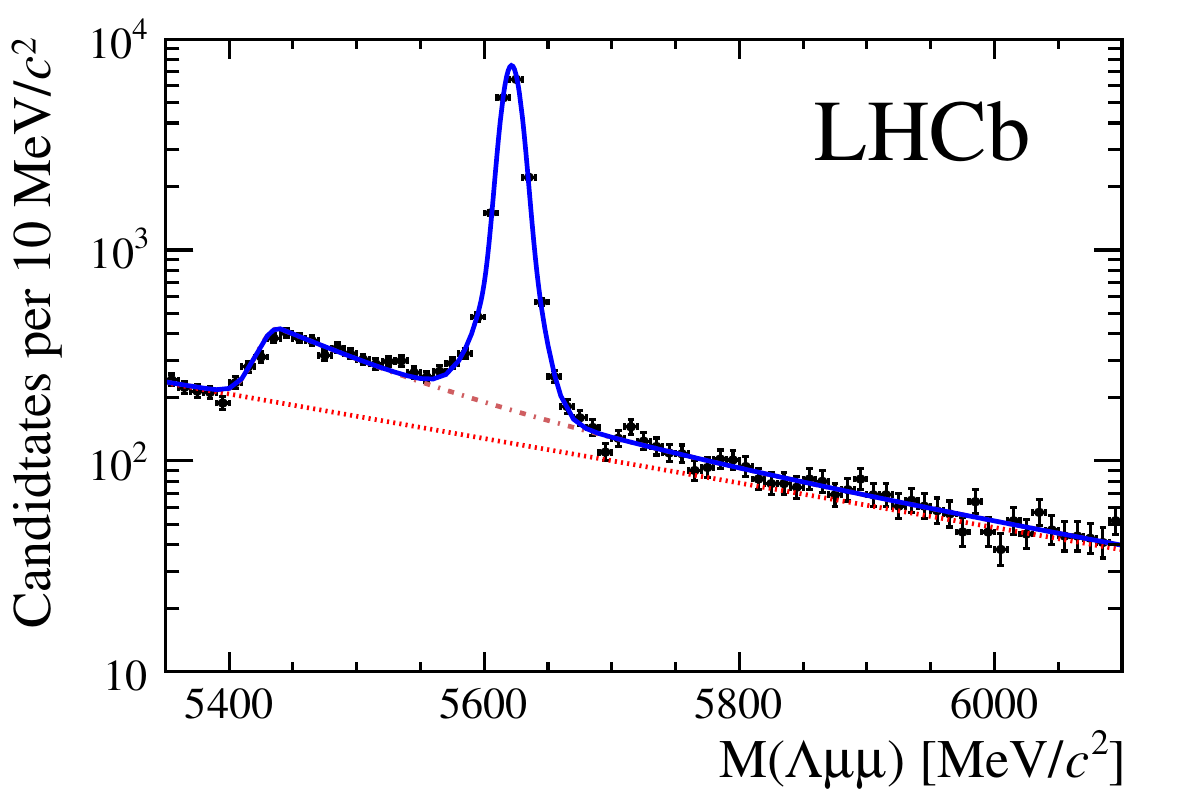}
\caption{\small Invariant mass distribution of the \decay{\Lb}{\jpsi\Lz}
  candidates selected with the neural network requirement used for the high-\qsq region.
  The (black) points show data, combining downstream
  and long candidates, and the solid (blue) line represents the
  overall fit function.  The dotted (red) line represents the combinatorial
  and the dash-dotted (brown) line the peaking background from
  \decay{\Bz}{\jpsi\KS} decays.}
\label{fig:totalFit}
\end{figure}

 The invariant mass distribution of the \decay{\Lb}{\jpsi\Lz}
 candidates selected with the high-\qsq requirements is shown in
 Fig.~\ref{fig:totalFit}, combining both long and downstream
 candidates.  The normalisation channel candidates are divided into
 four sub-samples: downstream and long events are fitted separately
 and each sample is selected with both the low-\qsq and high-\qsq
 requirements to normalise the corresponding \qsq regions in signal.
 The number of \decay{\Lb}{\jpsi\Lz} decays found in each case
 is given in Table~\ref{tab:jpsi_yield}.
\begin{table}[tbp]
\centering
\caption{Number of \decay{\Lb}{\jpsi\Lz} decays in the long and
  downstream categories found using the selection for low- and
  high-\qsq regions. Uncertainties shown are statistical only.}
\begin{tabular}{lcc}
Selection & $N_{\rm S}$ (long) & $N_{\rm S}$ (downstream)\\ \hline
high-\qsq	& $4313 \pm 70$	 	&  $11\,497 \pm 123$ \\
low-\qsq	& $3363 \pm 59$ 	&  $\phantom{0}\,7225 \pm 89\phantom{0}$  \\
 \hline
\end{tabular}
\label{tab:jpsi_yield}
\end{table}

  The fraction of peaking background events is larger in the
  downstream sample amounting to 28\,\% of the \decay{\Lb}{\jpsi\Lz} yield in the
  full fitted mass range, while in the sample of long candidates it
  constitutes about 4\,\%.

\begin{figure}[tbph!]
\centering \includegraphics[width=0.8\textwidth]{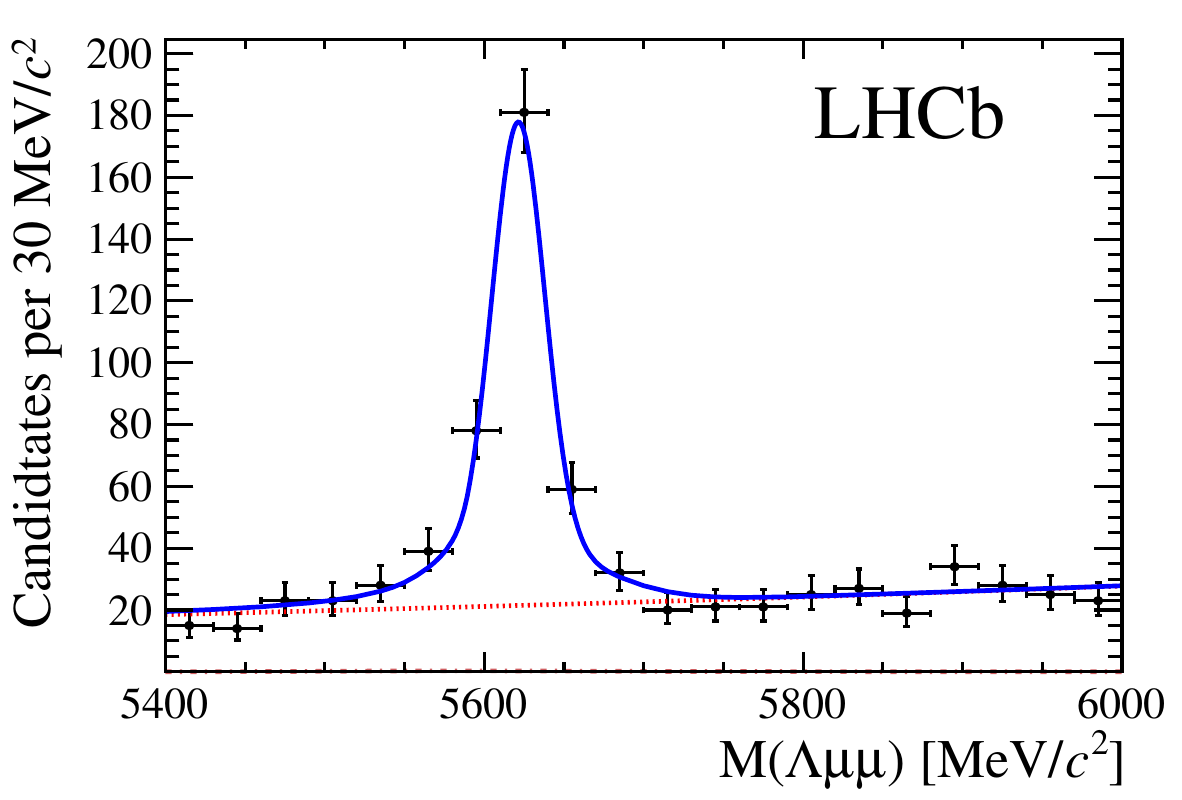}
\caption{\small Invariant mass distribution of the
  \decay{\Lb}{\Lz\mumu} candidates, integrated over the region  $15.0 < \qsq < 20.0$ \gevgevcccc 
  together with the fit function described in the text.  The points show data,
  the solid (blue) line is the overall fit function and the dotted
  (red) line represents the combinatorial background.
  The background component from \decay{\Bz}{\KS\mumu} decays, (brown)
  dashed line, is barely visibile due to the very low yield.}
\label{fig:totalFitRare}
\end{figure}

\begin{figure}[tbph]
\centering \includegraphics[width=0.64\textheight]{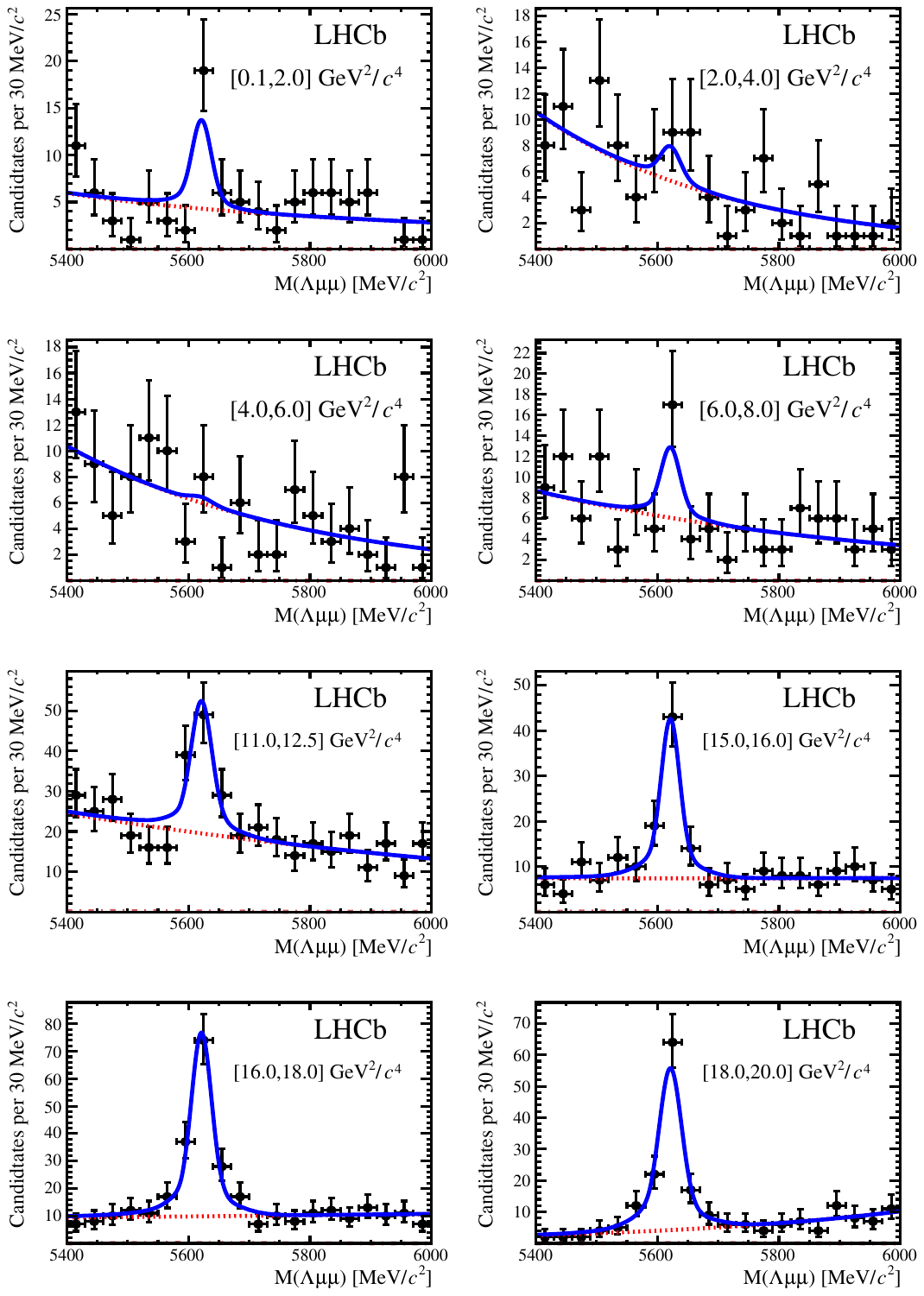}
\caption{\small Invariant mass distributions of \decay{\Lb}{\Lz\mumu}
candidates, in eight \qsq\ intervals, together with the
  fit function described in the text. The points show data, the
  solid (blue) line is the overall fit function and
  the dotted
  (red) line represents the combinatorial background component.}
\label{fig:differentialFit}
\end{figure}

  The invariant mass distributions for the \decay{\Lb}{\Lz\mumu}
  process, integrated over $15.0<\qsq<20.0$\gevgevcccc and in eight
  separate \qsq intervals, are shown in Figs.~\ref{fig:totalFitRare}
  and \ref{fig:differentialFit}.  The yields found in each \qsq
  interval are given in Table~\ref{tab:rareYields} together with their
  significances.  The statistical significance of the observed signal
  yields is evaluated as $\sqrt{2\Delta\ln{\mathcal{L}}}$, where
  $\Delta\ln{\mathcal{L}}$ is the change in the logarithm of the
  likelihood function when the signal component is excluded from the
  fit, relative to the nominal fit in which it is present.

\begin{table}[btph!]
\centering
\caption{\small Signal decay yields ($N_\mathrm{S}$) obtained from the
  mass fit to \decay{\Lb}{\Lz\mumu} candidates in each \qsq interval
  together with their statistical significances. 
  The yields are the sum of long and downstream categories with
  downstream decays comprising $\sim 80\,\%$ of the total yield.
  The $8-11$ and $12.5-15$ \gevgevcccc ~\qsq intervals are excluded
  from the study as they are dominated by decays via charmonium resonances.
  }
\label{tab:rareYields}
\begin{tabular}{ccc}
 $q^2$ interval [\gevgevcccc] & Total signal yield & Significance \\ \hline
0.1 -- 2.0    &  $16.0\pm5.3$            &  4.4 \\
2.0 -- 4.0    &  $\phantom{0}4.8\pm4.7$  &  1.2 \\
4.0 -- 6.0    &  $\phantom{0}0.9\pm2.3$  &  0.5 \\
6.0 -- 8.0    &  $11.4\pm5.3$            &  2.7 \\
11.0 -- 12.5  &  $\phantom{.0}60\pm12\phantom{.}$           &  6.5 \\
15.0 -- 16.0  &  $57\pm9$                &  8.7 \\
16.0 -- 18.0  &  $118\pm13$              &  13  \\
18.0 -- 20.0  &  $\phantom{.}100\pm11\phantom{.}$   &  14  \\
\hline
1.1 -- 6.0    &  $\phantom{0}9.4\pm6.3$  &  1.7 \\
15.0 -- 20.0  &  $276\pm20$              &  21  \\
\end{tabular}  
\end{table}


 \section{Relative efficiency}
\label{sec:efficiency}

 The measurement of the differential branching fraction of
 \decay{\Lb}{\Lz\mumu} relative to \decay{\Lb}{\jpsi\Lz} benefits from
 the cancellation of several potential sources of systematic
 uncertainty in the ratio of efficiencies, $\varepsilon^{\rm
   rel}=\etot(\decay{\Lb}{\Lz\mumu})/\etot(\decay{\Lb}{\jpsi\Lz})$.
 Due to the long lifetime of \Lz baryons, most of the candidates are
 reconstructed in the downstream category, with an overall efficiency
 of 0.20\,\%, while the typical efficiency is 0.05\,\% for long
 candidates.

 The efficiency of the PID is obtained from a data-driven
 method~\cite{LHCb-DP-2012-003} and found to be 98\,\% while all other
 efficiencies are evaluated using simulated data.  The models used for
 the simulation are summarised in Sec.~\ref{sec:Detector}.  The
 trigger efficiency is calculated using simulated data and increases
 from approximately 56\,\% to 86\,\% between the lowest and highest
 \qsq regions. An independent cross-check of the trigger efficiency is
 performed using a data-driven method.  This exploits the possibility
 of categorising a candidate \decay{\Lb}{\Lz\mumu} or
 \decay{\Lb}{\jpsi\Lz} decay in two ways depending on which tracks are
 directly responsible for its selection by the trigger: {``trigger on
   signal''} candidates, where the tracks responsible for the
 {hardware and software} trigger decisions are associated with the
 signal; and {``trigger independent of signal''} candidates, with a
 \Lb baryon reconstructed in either of these channels but where the
 trigger decision does not depend on any of their decay products.  As
 these two categories of event are not mutually exclusive, their
 overlap may be used to estimate the efficiency of the trigger
 selection using data.  Using \decay{\Lb}{\jpsi\Lz} candidates and
 calculating the ratio of yields that are classified as both trigger
 on signal and independent of signal, relative to those that are
 classified as trigger independent of signal, an efficiency of
 $(70\pm5)$\,\% is obtained, which is consistent with that of
 $(73.33\pm0.02)$\,\% computed from simulation.

\begin{figure}[tbp]
\centering \includegraphics[width=0.8\textwidth]{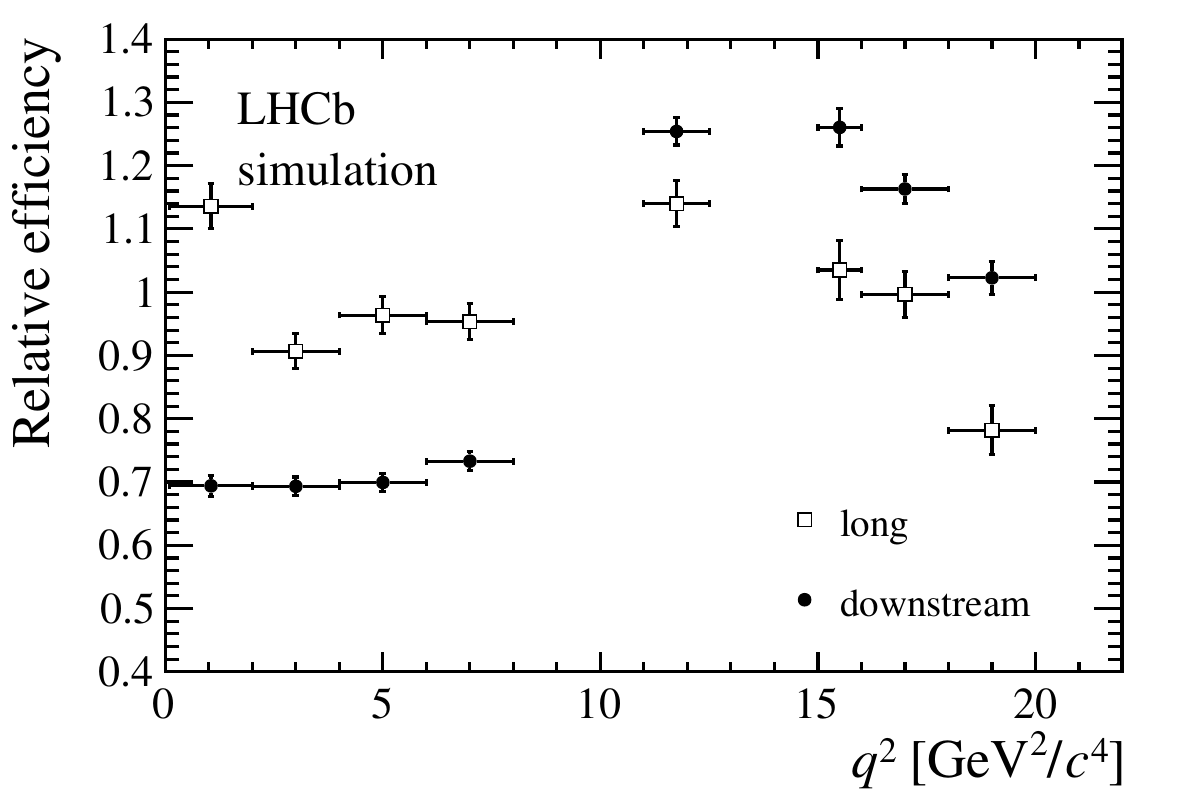}
\caption{\small
Total relative efficiency,
$\varepsilon_{\mathrm{rel}}$, between \decay{\Lb}{\Lz\mumu} and
\decay{\Lb}{\jpsi\Lz} decays.
The uncertainties are the combination of
both statistical and systematic components, and are dominated by the
latter.}
\label{fig:relativeTotalEfficiency}
\end{figure}

 The relative efficiency for the ratio of branching fractions in each
 \qsq interval, calculated from the absolute efficiencies described
 above, is shown in Fig.~\ref{fig:relativeTotalEfficiency}.
 The increase in efficiency as a function of increasing \qsq is
 dominated by two effects. Firstly, at low \qsq the muons have lower
 momenta and therefore have a lower probability of satisfying the
 trigger requirements.  Secondly, at low \qsq the \Lz baryon has a
 larger fraction of the \Lb momentum and is more likely to decay
 outside of the acceptance of the detector. 
 Separate selections are used for the low- and high-\qsq regions and,
 as can be seen in Fig.~\ref{fig:relativeTotalEfficiency}, the tighter
 neural network requirement used in the low-\qsq region has
 a stronger effect on downstream candidates.
 
 The uncertainties combine both
 statistical and systematic contributions (with the latter dominating)
 and include a small correlated uncertainty due to the use of a single
 simulated sample of \decay{\Lb}{\jpsi\Lz} decays as the normalisation channel
 for all \qsq\ intervals.  Systematic uncertainties associated with the
 efficiency calculation are described in detail in Sec.~\ref{sec:systemtics}.

\section{Systematic uncertainties on the branching fraction}
\label{sec:systemtics}
\subsection{Yields}
\label{sec:systematics_yields}
 Three sources of systematic uncertainty on the measured yields are
 considered for both the \decay{\Lb}{\jpsi\Lz} and the
 \decay{\Lb}{\Lz\mumu} decay modes: the shape of the signal PDF, the
 shape of the background PDF and the choice of the fixed parameters
 used in the fits to data.

 For both decays, the default signal PDF is replaced by the sum of two
 Gaussian functions. All parameters of the Gaussian functions are
 allowed to vary to take into account the effect of fixing
 parameters. The shape of the background function is changed by
 permitting the \KS\mumu peaking background yield, which is fixed to
 the value obtained from simulation the nominal fit, to vary.  For the
 resonant channel, the \jpsi\KS peaking background shape is changed by
 fixing the global shift to zero.  Finally, simulated experiments are
 performed using the default model, separately for each \qsq interval,
 generating the same number of events as observed in data.  Each
 distribution is fitted with the default model and the modified
 PDFs. The average deviation over the ensemble of simulated
 experiments is assigned as the systematic uncertainty.  The relative
 change in signal yield due to the choice of signal PDF varies between
 0.6\,\% and 4.6\,\% depending on \qsq, while the change due to the
 choice of background PDF is in the range between 1.1\,\% and
 2.5\,\%. The \qsq intervals that are most affected are those in which
 a smaller number of candidates is observed and therefore there are
 fewer constraints to restrict potentially different PDFs.  The
 systematic uncertainties on the yield in each \qsq interval are
 summarised in Table~\ref{tab:brsys}, where the total is the sum in
 quadrature of the individual components.

\subsection{Relative efficiencies}
 \label{sec:systematics_eff}
 The dominant systematic effect is that related to the current
 knowledge of the angular structure and the \qsq dependence of the
 decay channels.  The uncertainty due to the finite size of simulated
 samples is comparable to that from other sources. The total
 systematic uncertainties on the efficiencies, calculated as the sums
 in quadrature of the individual components described below,
 are summarised in Table~\ref{tab:brsys}.

\subsubsection{Decay structure and production polarisation}
 The main factors that affect the detection efficiencies are the
 angular structure of the decays and the production polarisation
 ($P_b$). Although these arise from different parts of the process,
 the efficiencies are linked and are therefore treated together.

 For the \decay{\Lb}{\Lz\mumu} decay, the impact of the limited
 knowledge of the production polarisation, $P_b$, is estimated by
 comparing the default efficiency, obtained in the unpolarised
 scenario, with those in which the polarisation is varied within its
 measured uncertainties, using the most recent LHCb measurement, $P_b
 = 0.06 \pm 0.09$\cite{LHCb-PAPER-2012-057}.  The larger of these
 differences is assigned as the systematic uncertainty from this
 source. This yields a $\sim 0.5\,\%$ uncertainty on the efficiency of
 downstream candidates and $\sim 1.2\,\%$ for long candidates.  No
 significant \qsq dependence is found.

 To assess the systematic uncertainty due to the limited knowledge of
 the decay structure, the efficiency corresponding to the default
 model \cite{Gutsche:2013pp,Aliev:2005np,Buras:1994dj} is compared to
 that of a model containing an alternative set of form factors based
 on a lattice QCD calculation \cite{Detmold:2012vy}. The larger of the
 full difference or the statistical precision is assigned as the
 systematic uncertainty.
 
 For the \decay{\Lb}{\jpsi\Lz} mode, the default angular distribution
 is based on that observed in Ref.  \cite{LHCb-PAPER-2012-057}. The
 angular distribution is determined by the production polarisation and
 four complex decay amplitudes. The central values from
 Ref.~\cite{LHCb-PAPER-2012-057} are used for the nominal result. To
 assess the sensitivity of the \decay{\Lb}{\jpsi\Lz} mode to the
 choice of decay model, the production polarisation and decay
 amplitudes are varied within their uncertainties, taking into account
 correlations.

 To assess the potential impact that physics beyond the SM might have
 on the detection efficiency, the $C_7$ and $C_9$ Wilson coefficients
 are modified by adding a non-SM contribution ($C_i \rightarrow C_i +
 C_i^{'}$). The $C_i^{'}$ added are inspired to maintain compatibility
 with the recent LHCb result for the $P'_5$ observable
 \cite{Descotes-Genon:2013wba} and indicate a change at the level of
 $\sim 7$\,\% in the 0.1--2.0 \qsq interval, and 2--3\,\% in other
 regions.  No systematic is assigned as a result of this study.

 \subsubsection{Reconstruction efficiency for the \Lz baryon}
 The \Lz baryon is reconstructed from either long or downstream
 tracks, and their relative proportions differ in data and simulation.
 This proportion does not depend significantly on \qsq and therefore
 possible effects cancel in the ratio with the normalisation channel.
 Furthermore, since the analysis is performed separately for long and
 downstream candidates, it is not necessary to assign a systematic
 uncertainty to account for a potential effect due to the different
 fractions of candidates of the two categories observed in data and
 simulation.  To allow for residual differences between data and
 simulation that do not cancel completely in the ratio between signal
 and normalisation modes, systematic uncertainties of 0.8\,\% and
 1.2\,\% are estimated for the low-\qsq and high-\qsq regions,
 respectively, using the same data-driven method as in
 Ref.~\cite{LHCb-PAPER-2014-006}.

 \subsubsection{Production kinematics and lifetime of the \Lb baryon}
  In \decay{\Lb}{\jpsi\Lz} decays a small difference is observed
  between data and simulation in the momentum and transverse momentum
  distributions of the \Lb baryon produced. Simulated data are
  reweighted to reproduce these distributions in data and the relative
  efficiencies are compared to those obtained using events that are
  not reweighted.  This effect is less than 0.1\,\%, which is
  negligible with respect to other sources.

  Finally, the \Lb baryon lifetime used throughout corresponds to the
  most recent \lhcb measurement, $1.479\pm0.019$\ps
  \cite{LHCb-PAPER-2014-003}.  The associated systematic uncertainty
  is estimated by varying the lifetime value by one standard deviation
  and negligible differences are found.

 \begin{table}[tbp]
 \centering
 \caption{Systematic uncertainties as a function of \qsq, assigned for
   yields and efficiencies. Values reported are the sums in quadrature
   of all contributions evaluated within each category. }
 \label{tab:brsys}
 \renewcommand{\arraystretch}{1.3}
 \begin{tabular}{ccc}
 $q^2$ interval [\gevgevcccc]  & Syst.\ on yields [\%] & Syst.\ on
  eff. [\%] \\ \hline

0.1 -- 2.0  	 	& 3.4  &	 $_{-3.6}^{+2.2}$ 	  \\
2.0 -- 4.0  	 	& 3.8  &	 $_{-4.1}^{+2.2}$ 	  \\
4.0 -- 6.0  	 	& 6.6  &	 $_{-14.3}^{+17.2}$   \\
6.0 -- 8.0  	 	& 2.0  &	 $_{-3.1}^{+2.1}$ 	  \\
11.0 -- 12.5  	 	& 3.2  &	 $_{-5.2}^{+3.7}$ 	  \\
15.0 -- 16.0  	 	& 2.8  &	 $_{-2.8}^{+3.1}$ 	  \\
16.0 -- 18.0  	 	& 1.4  &	 $_{-4.1}^{+3.0}$ 	  \\
18.0 -- 20.0  	 	& 2.5  &	 $_{-2.3}^{+3.9}$ 	  \\
\hline
1.1 -- 6.0  	 	& 4.2  &	 $_{-4.6}^{+2.2}$ 	  \\
15.0 -- 20.0       	& 1.0  &    $_{-2.9}^{+2.0}$      \\

\end{tabular}
\end{table}

\section{Differential branching fraction}
\begin{figure}[tbph]
 \centering
 \includegraphics[width=0.8\textwidth]{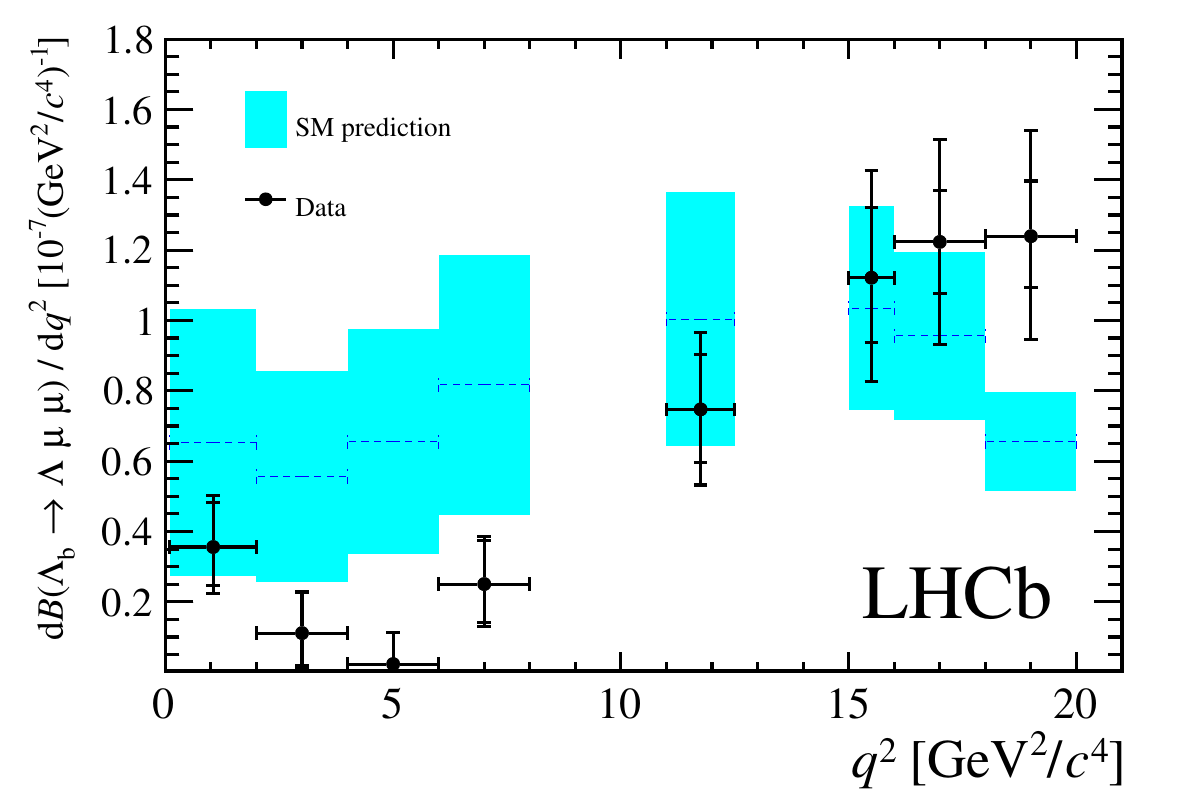}
 \protect\caption{ Measured \protect\decay{\Lb}{\Lz\mumu} branching
   fraction as a function of \qsq with the predictions of the SM
   \cite{Detmold:2012vy} superimposed.  The inner error bars on data
   points represent the total uncertainty on the relative branching
   fraction (statistical and systematic); the outer error bar also
   includes the uncertainties from the branching fraction of the
   normalisation mode.}  \protect\label{fig:mass_fit_smallbins}
 \end{figure}

 The values for the absolute branching fraction of the
 \decay{\Lb}{\Lz\mumu} decay, obtained by multiplying the relative
 branching fraction by the absolute branching fraction of the
 normalisation channel,
 $\BF(\decay{\Lb}{\jpsi\Lz})=(6.3\pm1.3)\times10^{-4}$~\cite{Agashe:2014kda},
 are given in Fig.~\ref{fig:mass_fit_smallbins} and summarised in
 Table~\ref{tab:AbsBR}, where the SM predictions are obtained from
 Ref.~\cite{Detmold:2012vy}.  The relative branching fractions are
 given in the Appendix.

 Evidence for signal is found in the \qsq region between the
 charmonium resonances and in the interval $0.1 < \qsq < 2.0$
 \gevgevcccc, where an increased yield is expected due to the
 proximity of the photon pole.  The uncertainty on the branching
 fraction is dominated by the precision of the branching fraction for
 the normalisation channel, while the uncertainty on the relative
 branching fraction is dominated by the size of the data sample
 available.  The data are consistent with the theoretical predictions
 in the high-\qsq region but lie below the predictions in the low-\qsq
 region.

\begin{table}[tbph]
\centering
\renewcommand{\arraystretch}{1.2}
\caption{Measured differential branching fraction of
  \decay{\Lb}{\Lz\mumu}, where the uncertainties are statistical, systematic and
 due to the uncertainty on the normalisation mode, \decay{\Lb}{\jpsi\Lz}, respectively.}
\begin{tabular}{ccccccc}
  \qsq interval  [\gevgevcccc] & &\multicolumn{5}{c}{$\deriv \BF(\decay{\Lb}{\Lz\mumu})/\deriv\qsq \cdot 10^{-7} [(\gevgevcccc)^{-1}]$} \\
\hline
0.1 -- 2.0    &    &0.36  &  $^{+\,0.12}_{-\,0.11}$   & $^{+\,0.02}_{-\,0.02}$ & $\pm\,0.07$ \\
2.0 -- 4.0    &    &0.11  &  $^{+\,0.12}_{-\,0.09}$   & $^{+\,0.01}_{-\,0.01}$ & $\pm\,0.02$ \\
4.0 -- 6.0    &    &0.02  &  $^{+\,0.09}_{-\,0.00}$   & $^{+\,0.01}_{-\,0.01}$ & $\pm\,0.01$ \\
6.0 -- 8.0    &    &0.25  &  $^{+\,0.12}_{-\,0.11}$   & $^{+\,0.01}_{-\,0.01}$ & $\pm\,0.05$ \\

11.0 -- 12.5  &    &0.75  &  $^{+\,0.15}_{-\,0.14}$   & $^{+\,0.03}_{-\,0.05}$ & $\pm\,0.15$ \\
15.0 -- 16.0  &    &1.12  &  $^{+\,0.19}_{-\,0.18}$   & $^{+\,0.05}_{-\,0.05}$ & $\pm\,0.23$ \\
16.0 -- 18.0  &    &1.22  &  $^{+\,0.14}_{-\,0.14}$   & $^{+\,0.03}_{-\,0.06}$ & $\pm\,0.25$ \\
18.0 -- 20.0  &    &1.24  &  $^{+\,0.14}_{-\,0.14}$   & $^{+\,0.06}_{-\,0.05}$ & $\pm\,0.26$ \\

\hline
1.1 -- 6.0    &    &0.09  &  $^{+\,0.06}_{-\,0.05}$   & $^{+\,0.01}_{-\,0.01}$ & $\pm\,0.02$ \\
15.0 -- 20.0  &    &1.20  &  $^{+\,0.09}_{-\,0.09}$   & $^{+\,0.02}_{-\,0.04}$ & $\pm\,0.25$ \\
 \end{tabular}
\label{tab:AbsBR}
\end{table}

\section{Angular analysis}
The forward-backward asymmetries of both the dimuon system, $A_{\rm
  FB}^\ell$, and of the $p\pi$ system, $A_{\rm FB}^h$, are defined as
\begin{align}
A_{\rm FB}^i(\qsq)&=\frac{\int_0^1 \frac{\deriv^2\Gamma}{\deriv\qsq\,\deriv\!\cos\theta_i} \deriv\!\cos\theta_i-
               \int^0_{-1} \frac{\deriv^2\Gamma}{\deriv\qsq\,\deriv\!\cos\theta_i} \deriv\!\cos\theta_i}{\deriv\Gamma / \deriv \qsq},
\label{eq:afbTh}
\end{align}
\noindent
where $\deriv^2\Gamma/\deriv q^2\,\deriv\!\cos\theta_i$ is the
two-dimensional differential rate and $\deriv\Gamma/\deriv q^2$ is the
rate integrated over the corresponding angles.  The observables are
determined by a fit to one-dimensional angular distributions as a
function of $\cos \theta_\ell$, the angle between the positive
(negative) muon direction and the dimuon system direction in the
\Lb(\Lbbar) rest frame, and $\cos \theta_h$, which is defined as the
angle between the proton and the \Lz baryon directions, also in the
\Lb rest frame. The differential rate as a function of $\cos
\theta_\ell$ is described by the function
\begin{align}
\frac{\deriv^2\Gamma(\Lambda_{b}\to \Lambda \,\ell^{+}\ell^{-})}{\deriv\qsq\,\deriv\!\cos\theta_\ell}=
\frac{\deriv\Gamma}{\deriv\qsq}&\left[  \frac{3}{8}\left(1+\cos^2\theta_\ell\right)(1-f_{\rm L})+A_{\rm FB}^\ell\cos\theta_\ell +
   \frac{3}{4}f_{\rm L} \sin^2\theta_\ell \right], 
\label{eq:afbLTh}
\end{align}
where $f_{\rm L}$ is the fraction of longitudinally polarised dimuons.
 The rate as a function of $\cos \theta_h$ has the form
\begin{equation}
\frac{\deriv^2\Gamma(\Lambda_{b}\to \Lambda(\to p \pi^{-})\ell^{+}\ell^{-})}
     {\deriv\qsq\,\deriv\!\cos\theta_h} 
={\BF}(\Lambda \to p\pi^{-})
 \frac{\deriv\Gamma(\Lambda_b \to \Lambda\, \ell^{+}\ell^{-})}{\deriv \qsq}\frac{1}{2}
\Big(1+2A_{\rm FB}^h\cos\theta_h\Big) \,.
\label{eq:afbBTh}
\end{equation}
\noindent
These expressions assume that \Lb baryons are produced unpolarised,
which is in agreement with the measured production polarisation at
\lhcb~\cite{LHCb-PAPER-2012-057}.

\begin{figure}[tbp]
\centering
\includegraphics[width=0.45\textwidth]{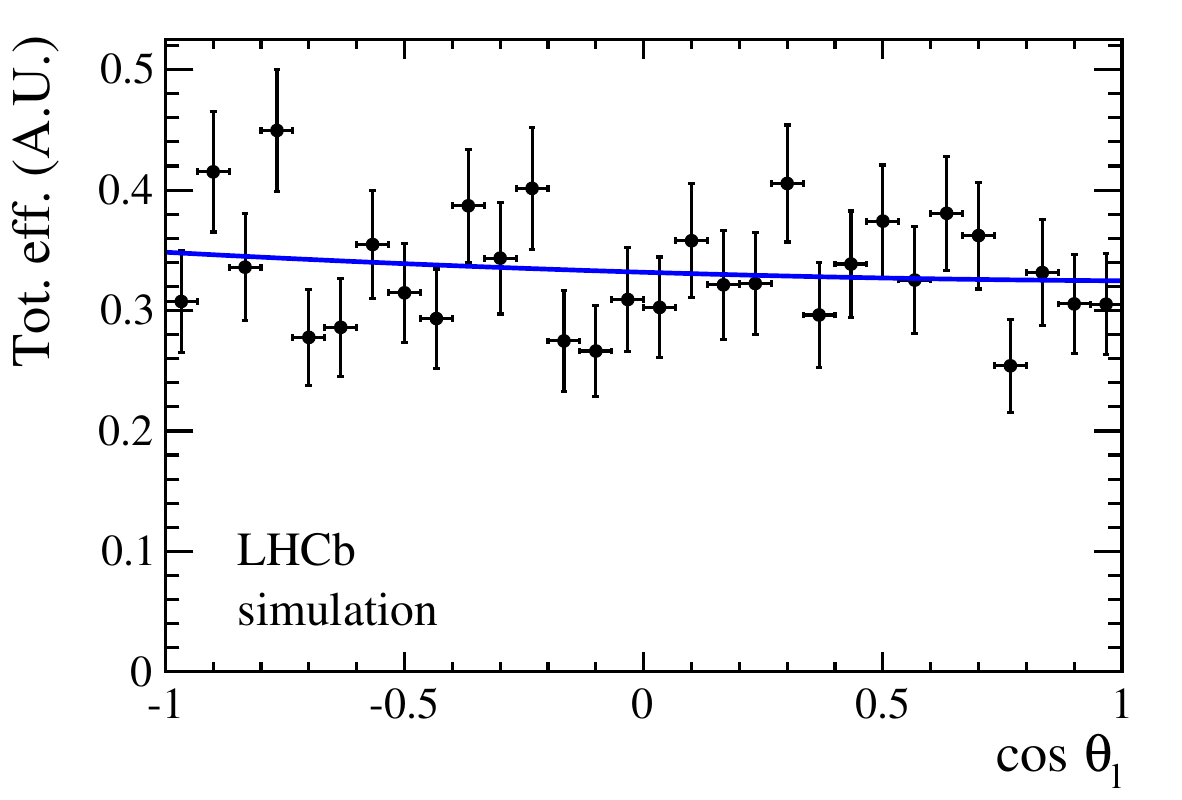}
\includegraphics[width=0.45\textwidth]{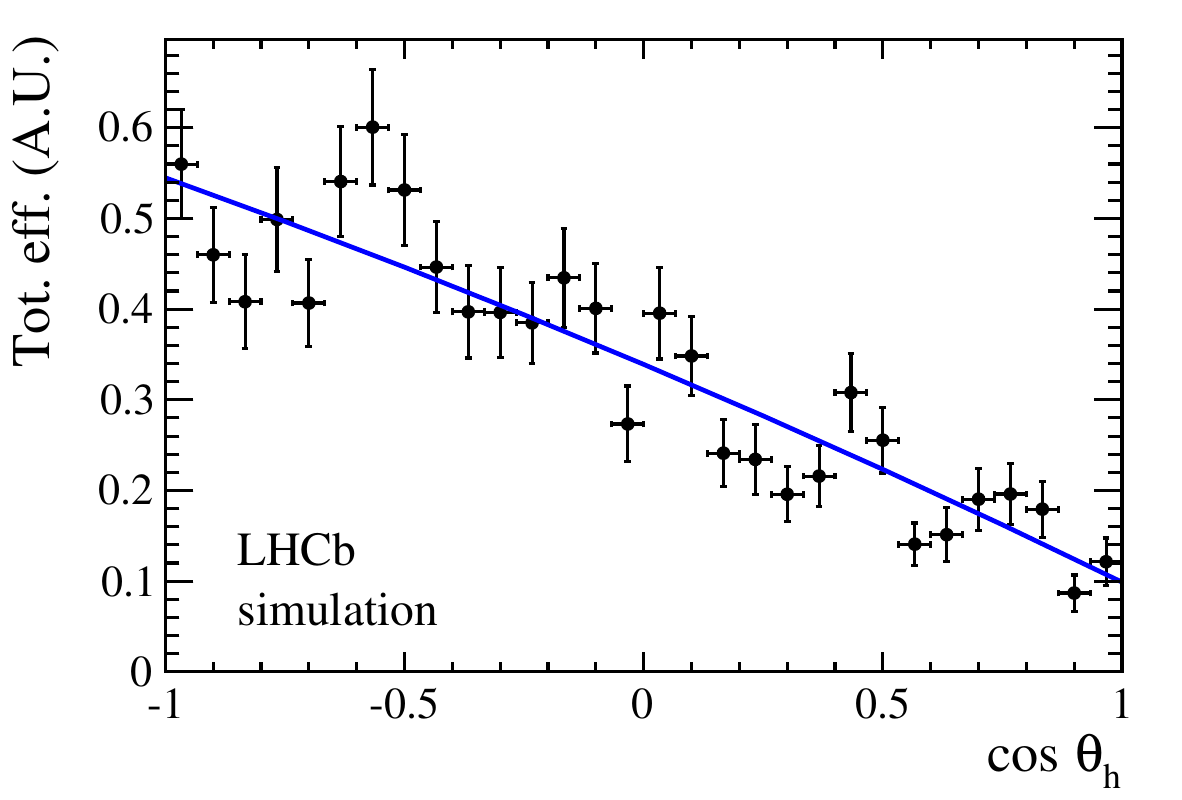}
\includegraphics[width=0.45\textwidth]{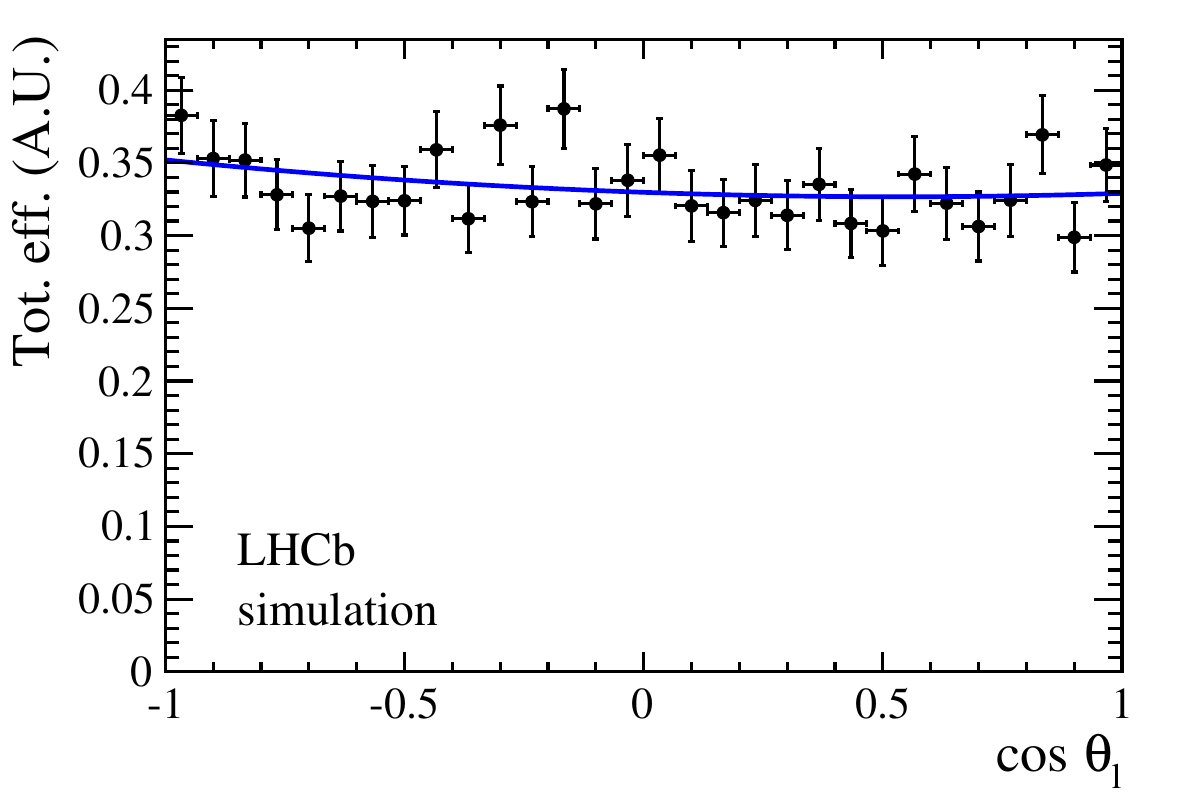}
\includegraphics[width=0.45\textwidth]{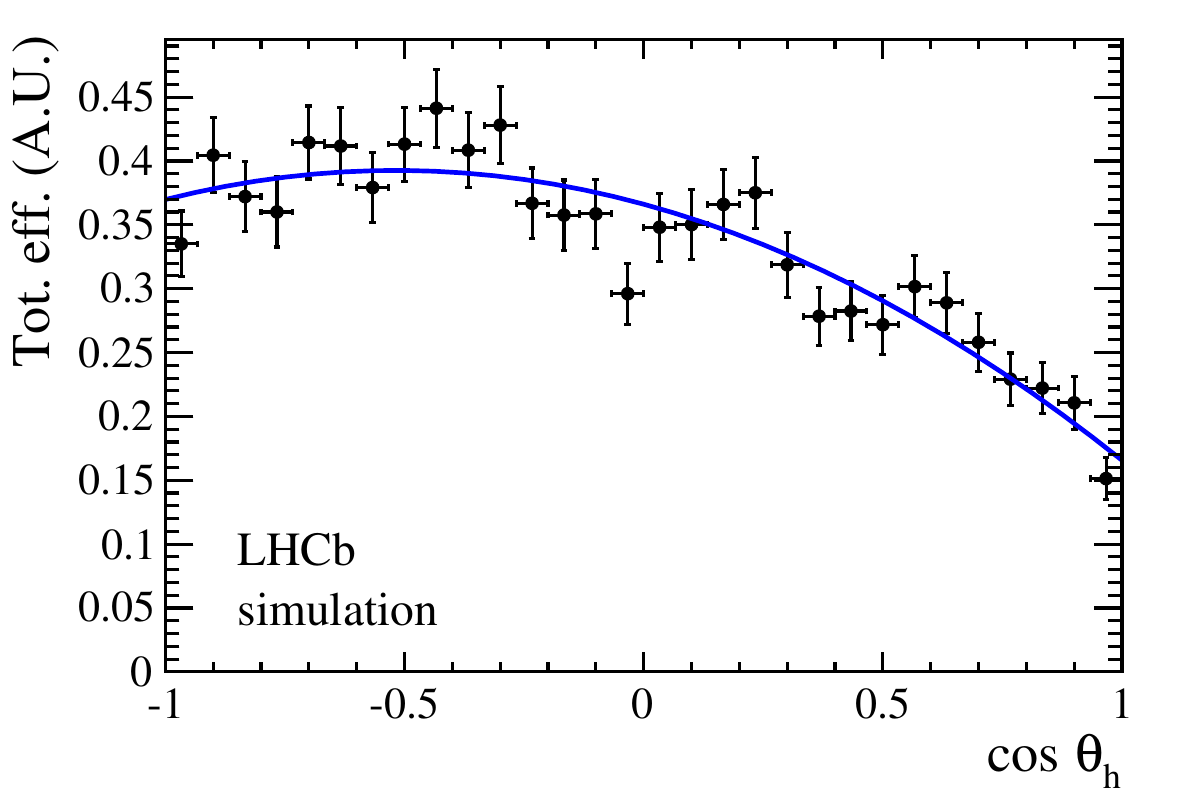}

\caption{Angular efficiencies as a function of (left)
  $\cos\theta_\ell$ and (right) $\cos\theta_h$ for (upper) long and
  (lower) downstream candidates, in the interval $15 < \qsq < 20$
  \gevgevcccc, obtained using simulated events.  The (blue) line shows
  the fit that is used to model the angular acceptance in the fit to
  data. }
\label{fig:AngEff}
\end{figure}

The forward-backward asymmetries are measured in data using unbinned
maximum likelihood fits.  The signal PDF consists of a theoretical
shape, given by Eqs.~\ref{eq:afbLTh} and \ref{eq:afbBTh}, multiplied
by an acceptance function.  Selection requirements on the minimum
momentum of the muons may distort the $\cos \theta_\ell$ distribution
by removing candidates with extreme values of $\cos
\theta_\ell$. Similarly, the impact parameter requirements affect
$\cos \theta_h$ as very forward hadrons tend to have smaller impact
parameter values.  The angular efficiency is parametrised using a
second-order polynomial and determined separately for downstream and
long candidates by fitting simulated events, with an independent set
of parameters obtained for each \qsq interval. These parameters are
fixed in the fits to data.  The acceptances are shown in
Fig.~\ref{fig:AngEff} as a function of $\cos\theta_h$ and
$\cos\theta_\ell$ in the $15 < \qsq <20$ \gevgevcccc interval for each
candidate category.

The background shape is parametrised by the product of a linear
function and the signal efficiency, with the value of the slope
determined by fitting candidates in the upper mass sideband,
$m(\Lz\mumu) > 5700$ \mevcc.  To limit systematic effects due to
uncertainties in the background parametrisation, an invariant mass
range that is dominated by signal events is used: $5580 < m(\Lz\mumu)
< 5660 $ \mevcc.  The ratio of signal to background events in this
region is obtained by performing a fit to the invariant mass
distribution in a wider mass interval.

\begin{figure}[tbp]
\centering
\includegraphics[width=0.49\textwidth]{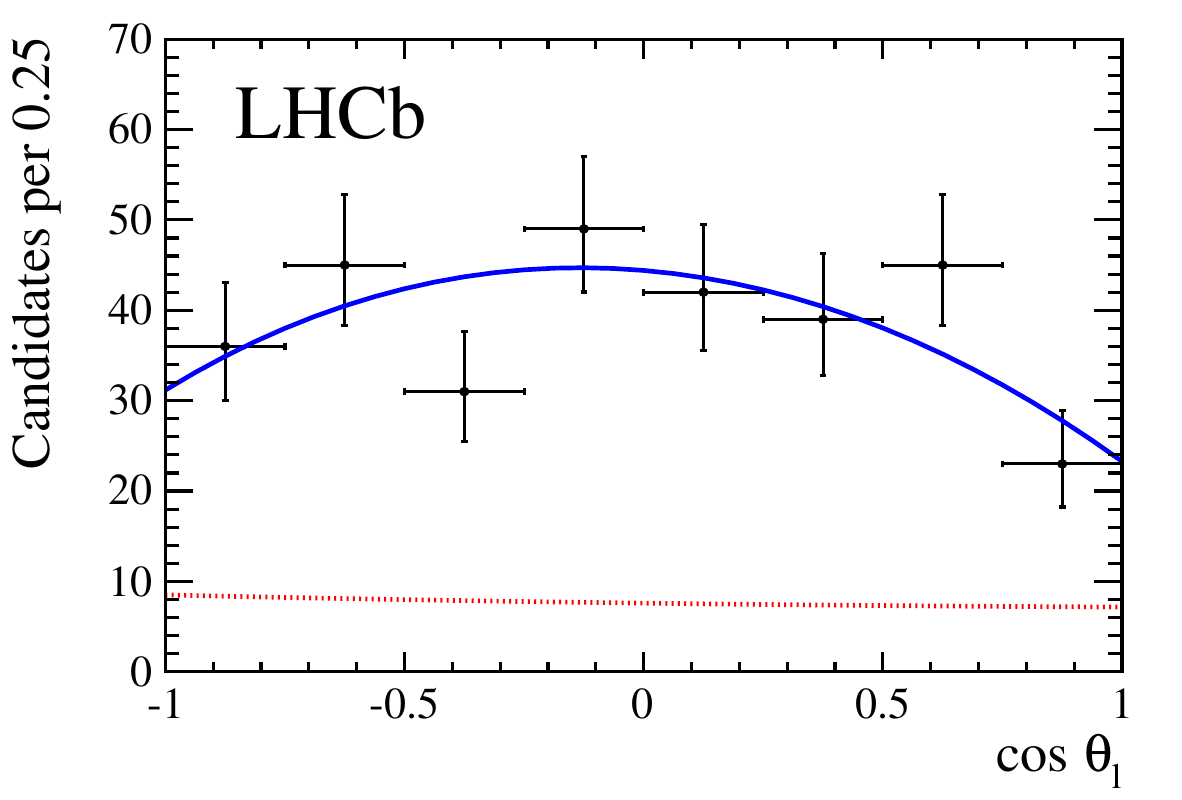}
\includegraphics[width=0.49\textwidth]{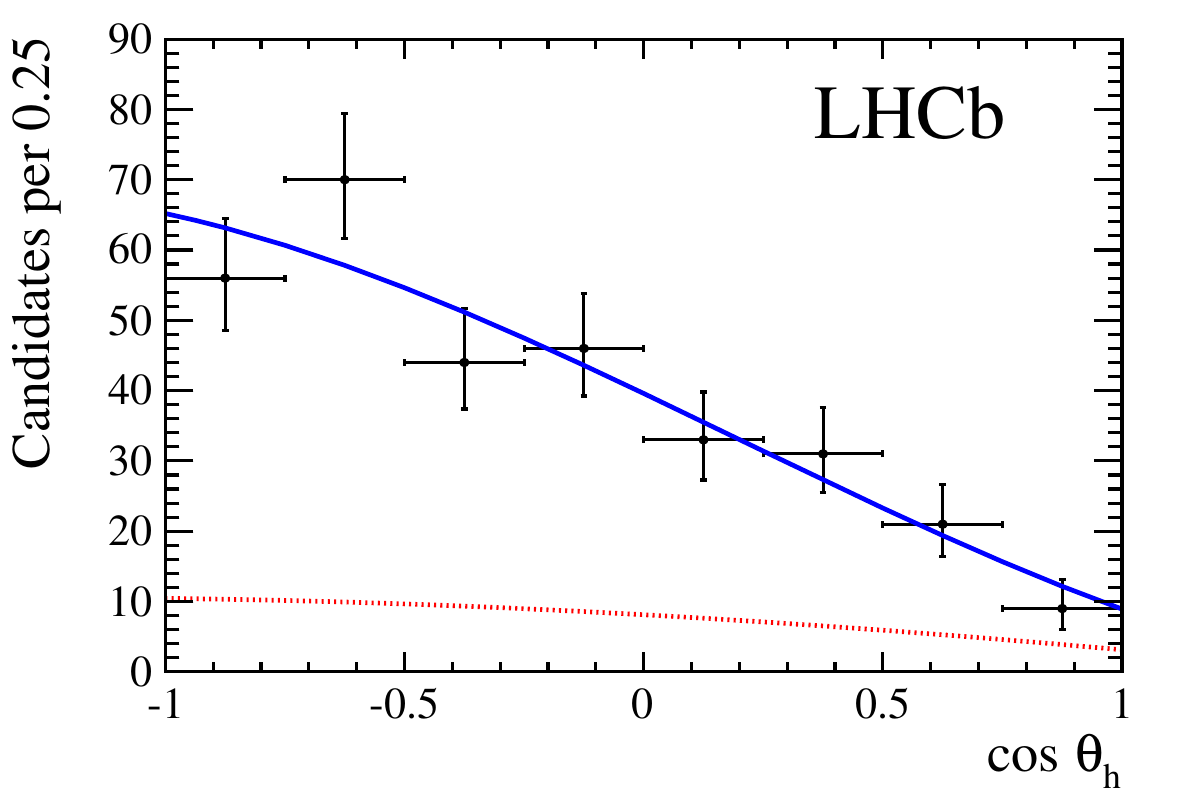}
\caption{Angular distributions as a function of (left) $\cos
  \theta_\ell$ and (right) $\cos \theta_h$, for candidates in the
  integrated $15 < \qsq <20$ \gevgevcccc interval with the overall fit
  function overlaid (solid blue). The (red) dotted line represents the
  combinatorial background.}
\label{fig:AngDistrib}
\end{figure}

The angular fit is performed simultaneously for the samples of downstream and
long candidates, 
using separate acceptance and background functions for the two
categories while keeping the angular observables as shared parameters.
Angular distributions are shown in Fig.~\ref{fig:AngDistrib} where the
two candidate categories are combined. 

\section{Systematic uncertainties on angular observables}
\subsection{Angular correlations}
 To derive Eqs.~\ref{eq:afbLTh} and \ref{eq:afbBTh}, a uniform
 efficiency is assumed. However, non-uniformity is observed,
 especially as a function of $\cos\theta_h$ (see
 Fig.~\ref{fig:AngEff}).  Therefore, while integrating over the full
 angular distribution, terms that would cancel with constant
 efficiency may remain and generate a bias in the measurement of these
 observables. To assess the impact of this potential bias, simulated
 experiments are generated in a two-dimensional ($\cos
 \theta_\ell$,$\cos \theta_h$) space according to the theoretical
 distribution multiplied by a two-dimensional efficiency histogram.
 Projections are then made and are fitted with the default
 one-dimensional efficiency functions. The average deviations from the
 generated parameters are assigned as systematic uncertainties. The
 magnitudes of these are found to be $-0.032$ for $A_{\rm FB}^\ell$,
 $0.013$ for $A_{\rm FB}^h$ and $0.028$ for $f_{\rm L}$, independently
 of \qsq.  In most \qsq intervals this is the dominant source of
 systematic uncertainty.

\subsection{Resolution}
 Resolution effects may induce an asymmetric migration of events
 between bins and therefore generate a bias in the measured value of
 the forward-backward asymmetries.  To study this systematic effect, a
 map of the angular resolution function is created using simulated
 events by comparing reconstructed quantities with those in the
 absence of resolution effects.  Simulated experiments are then
 generated according to the measured angular distributions and smeared
 using the angular resolution maps.  The simulated events, before and
 after smearing by the angular resolution function, are fitted with
 the default PDF.  The average deviations from the default values are
 assigned as systematic uncertainties.  These are larger for the
 $A_{\rm FB}^h$ observable because the resolution is poorer for
 $\cos\theta_h$ and the distribution is more asymmetric, yielding a
 net migration effect.  The uncertainties from this source are in the
 ranges $[0.011,0.016]$ for $A_{\rm FB}^\ell$, $[-0.001,-0.007]$ for
 $A_{\rm FB}^h$ and $[0.002,0.008]$ for $f_{\rm L}$, depending on
 \qsq.

\subsection{Angular acceptance}
 An imprecise determination of the efficiency due to data-simulation
 discrepancies could bias the $A_{\rm FB}$ measurement.  To estimate
 the potential impact arising from this source, the kinematic
 reweighting described in Sec.~\ref{sec:systematics_eff} is removed
 from the simulation.  Simulated samples are fitted using the same
 theoretical PDF multiplied by the efficiency function obtained with
 and without kinematical reweighting.  The average biases evaluated
 from simulated experiments are assigned as systematic uncertainties.
 These are larger for sparsely populated \qsq intervals and vary in
 the intervals $[0.009,0.016]$ for $A_{\rm FB}^\ell$, $[0.001,0.007]$
 for $A_{\rm FB}^h$ and $[0.002,0.044]$ for $f_{\rm L}$, depending on
 \qsq.

 The effect of the limited knowledge of the \Lb polarisation is
 investigated by varying the polarisation within its measured
 uncertainties, in the same way as for the branching fraction
 measurement. No significant effect is found and therefore no
 contribution is assigned.

\subsection{Background parametrisation}
\label{sec:bkgShapeSys}
As there is ambiguity in the choice of parametrisation for the
background model, in particular for regions with low statistical
significance in data, simulated experiments are generated from a PDF
corresponding to the best fit to data, for each \qsq interval. Each
simulated sample is fitted with two models: the nominal fit model,
consisting of the product of a linear function and the signal
efficiency, and an alternative model formed from a constant function
multiplied by the efficiency shape.  The average deviations are taken
as systematic uncertainties.  These are in the ranges $[0.003,0.045]$
for $A_{\rm FB}^\ell$, $[0.017,0.053]$ for $A_{\rm FB}^h$ and
$[0.014,0.049]$ for $f_{\rm L}$, depending on \qsq.

\section{Results of the angular analysis}
 The angular analysis is performed using the same \qsq intervals as
 those used in the branching fraction measurement.  Results are
 reported for each \qsq interval in which the statistical significance
 of the signal is at least three standard deviations. This includes
 all of the \qsq intervals above the \jpsi resonance and the lowest
 \qsq bin.

\begin{table}[tbp]
\centering
\caption{Measured values of leptonic and hadronic angular observables,
  where the first uncertainties are statistical and the second
  systematic.}
\label{tab:afbresults}
\renewcommand{\arraystretch}{1.2}
\begin{tabular}{c|ccc}
 \qsq interval  [\gevgevcccc]   &            $A_{\rm FB}^\ell$      &       $f_{\rm L}$ 						&  $A_{\rm FB}^h$                    \\ \hline

0.1 -- 2.0   & $\phantom{-\,}0.37 \; ^{+\;0.37}_{-\;0.48} \,\pm\, 0.03$  	&   $0.56 \; ^{+\;0.23}_{-\;0.56}\,\pm\, 0.08$ 		& $-\;0.12 \; ^{+\;0.31}_{-\;0.28}\,\pm\, 0.15$	\\
11.0 -- 12.5 & $\phantom{-\,}0.01 \; ^{+\;0.19}_{-\;0.18} \,\pm\, 0.06$  	&   $0.40 \; ^{+\;0.37}_{-\;0.36}\,\pm\, 0.06$		& $-\;0.50 \; ^{+\;0.10}_{-\;0.00}\,\pm\, 0.04$	 \\
15.0 -- 16.0 & $-\,0.10 \; ^{+\;0.18}_{-\;0.16} \,\pm\, 0.03$  			&   $0.49 \; ^{+\;0.30}_{-\;0.30} \,\pm\, 0.05$ 	& $-\;0.19 \; ^{+\;0.14}_{-\;0.16}\,\pm\, 0.03$	\\	
16.0 -- 18.0 & $-\,0.07 \; ^{+\;0.13}_{-\;0.12} \,\pm\, 0.04$  			&   $0.68 \; ^{+\;0.15}_{-\;0.21} \,\pm\, 0.05$ 	& $-\;0.44 \; ^{+\;0.10}_{-\;0.05}\,\pm\, 0.03$	\\
18.0 -- 20.0 & $\phantom{-\,}0.01 \; ^{+\;0.15}_{-\;0.14} \,\pm\; 0.04$  	&   $0.62 \; ^{+\;0.24}_{-\;0.27}\,\pm\, 0.04$ 		& $-\;0.13 \; ^{+\;0.09}_{-\;0.12}\,\pm\, 0.03$	\\ \hline
15.0 -- 20.0 & $-\,0.05 \; ^{+\;0.09}_{-\;0.09} \,\pm\, 0.03$  			&   $0.61 \; ^{+\;0.11}_{-\;0.14} \,\pm\, 0.03$ 	& $-\;0.29 \; ^{+\;0.07}_{-\;0.07}\,\pm\, 0.03$	\\
\end{tabular}
\end{table}

\begin{figure}[ptb]
\centering
\includegraphics[width=0.49\textwidth]{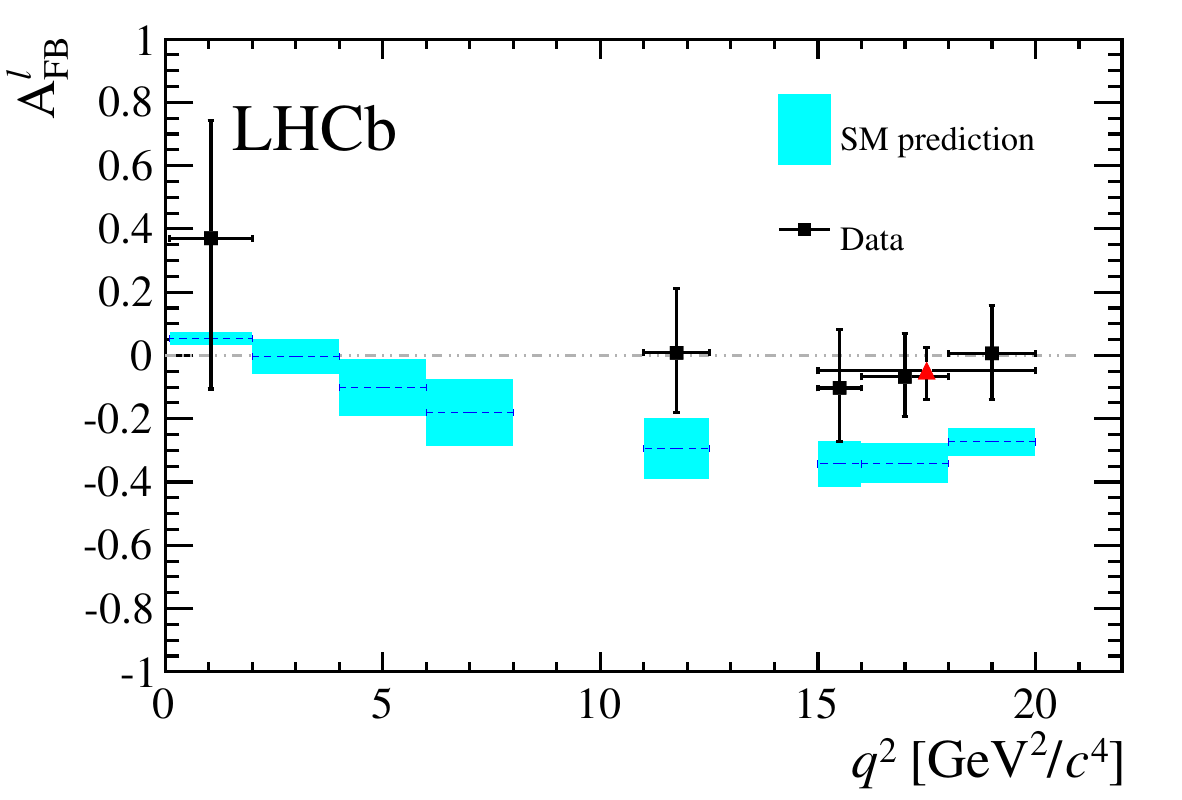}
\includegraphics[width=0.49\textwidth]{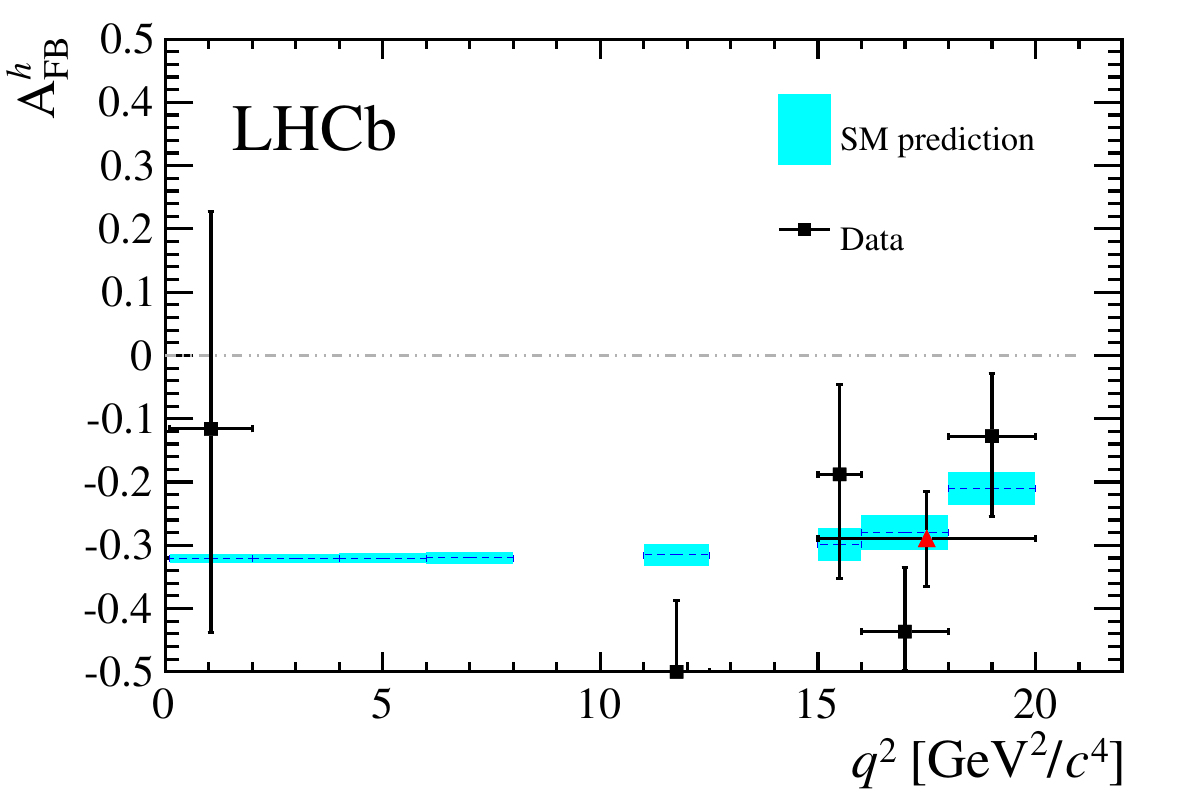}
\caption{Measured values of (left) the leptonic and (right) the hadronic
  forward-backward asymmetries in bins of \qsq.
  Data points are only shown for \qsq intervals where a statistically
  significant signal yield is found, see text for details.
  The (red) triangle represents the values for the $15 < \qsq < 20$ \gevgevcccc
  interval. Standard Model predictions are obtained from Ref.~\cite{Meinel:2014wua}.}
\label{fig:Afb_results}
\end{figure}

\begin{figure}[pbt]
\centering
\includegraphics[width=0.8\textwidth]{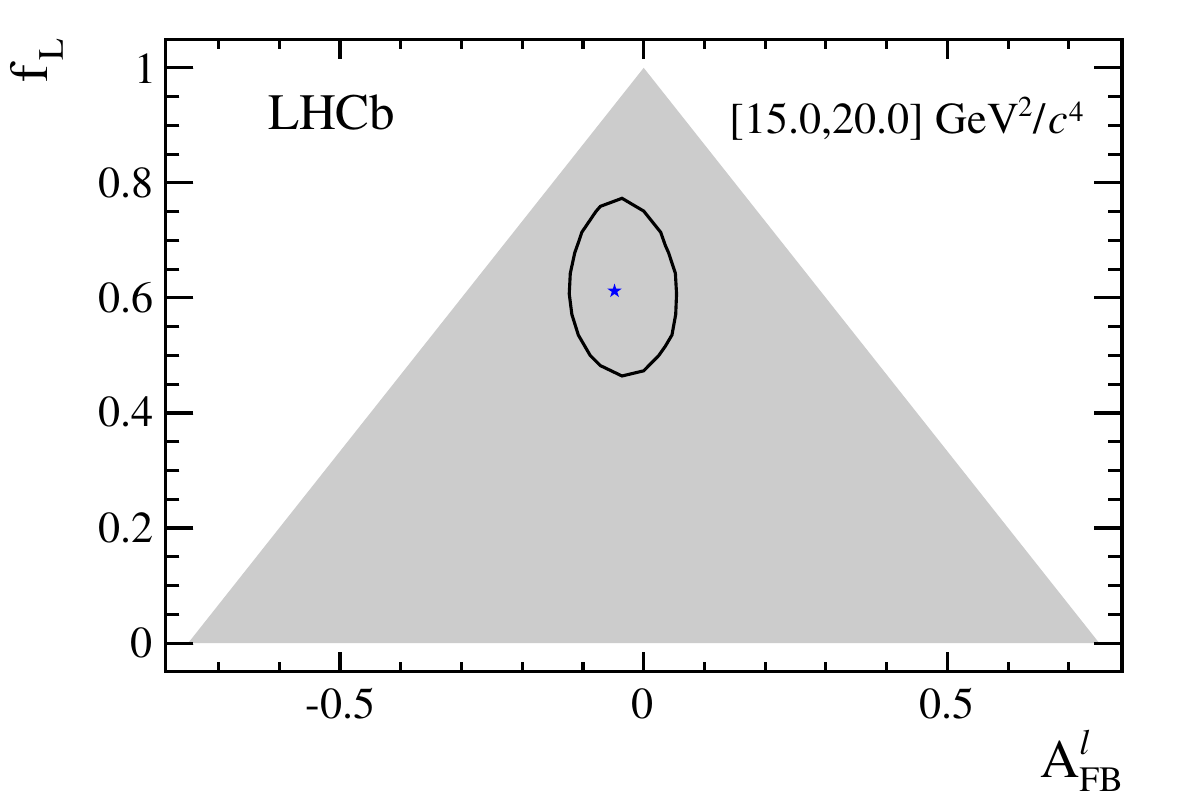}
\caption{Two-dimensional 68\,\% CL region (black) as a
  function of $A_{\rm FB}^\ell$ and $f_{\rm L}$.  The shaded area
  represents the region where the PDF is positive over the complete $\cos
  \theta_{\ell}$ range. The best fit point is given by the (blue) star. }
\label{fig:contour_highq2}
\end{figure}

The measured values of the leptonic and hadronic forward-backward
asymmetries, $A_{\rm FB}^\ell$\footnote{During preparation of update mistake in the analysis was identified, which changes the meaning of the measured quantity. Please see appendix \ref{sec:erratum} for details.} and $A_{\rm FB}^h$, and the $f_{\rm L}$
observable are summarised in Table~\ref{tab:afbresults}, with the
asymmetries shown in Fig.~\ref{fig:Afb_results}. The statistical
uncertainties are obtained using the likelihood-ratio ordering
method\cite{Feldman:1997qc} where only one of the two observables at a
time is treated as the parameter of interest.  In this analysis
nuisance parameters were accounted for using the plug-in
method~\cite{woodroofe}.  In Fig.~\ref{fig:contour_highq2} the
statistical uncertainties on $A_{\rm FB}^\ell$ and $f_{\rm L}$ are
also reported (for the interval $15 < \qsq < 20$ \gevgevcccc) as a
two-dimensional 68\;\% confidence level (CL) region, where the
likelihood-ratio ordering method is applied by varying both
observables and therefore taking correlations into account.
Confidence regions for the other \qsq intervals are shown in
Fig.~\ref{fig:contours}, see Appendix.

\section{Conclusions}
  A measurement of the differential branching fraction of the
  \decay{\Lb}{\Lz\mumu} decay is performed using data, corresponding
  to an integrated luminosity of 3.0\invfb, recorded by the \lhcb
  detector at centre-of-mass energies of 7 and 8\tev. Signal is
  observed for the first time at a significance of more than three
  standard deviations in two \qsq intervals: $0.1 < \qsq < 2.0$
  \gevgevcccc, close to the photon pole, and between the charmonium
  resonances. No significant signal is observed in the $1.1 < \qsq <
  6.0$ \gevgevcccc range. The uncertainties of the measurements in the
  region $15 < \qsq < 20$ \gevgevcccc are reduced by a factor of
  approximately three relative to previous \lhcb
  measurements~\cite{LHCB-PAPER-2013-025}.  The improvements in the
  results, which supersede those of Ref.~\cite{LHCB-PAPER-2013-025},
  are due to the larger data sample size and a better control of
  systematic uncertainties.  The measurements are compatible with the
  predictions of the Standard Model in the high-\qsq region and lie
  below the~predictions~in~the~low-\qsq~region.

  The first measurement of angular observables for the
  \decay{\Lb}{\Lz\mumu} decay is reported, in the form of two
  forward-backward asymmetries, in the dimuon and $p\pi$ systems and
  the fraction of longitudinally polarised dimuons.  The measurements
  of the $A_{\rm FB}^h$ observable are in good agreement with the
  predictions of the SM, while for the $A_{\rm FB}^\ell$ observable
  measurements are consistently above the prediction.

\clearpage
\appendix
\section{Appendix}
\label{app:A}

 The measured values of the 
 branching fraction of the \decay{\Lb}{\Lz\mumu} decay normalised to
 \decay{\Lb}{\jpsi\Lz} decays are given in Table~\ref{tab:RelBR},
 where the statistical and total systematic uncertainties are shown
 separately. 

\begin{table}[th]
\centering
\renewcommand{\arraystretch}{1.2}
\caption{Differential branching fraction of the \decay{\Lb}{\Lz\mumu}
  decay relative to \decay{\Lb}{\jpsi\Lz} decays,
 where the uncertainties are statistical and systematic, respectively.}
\begin{tabular}{cccccc}
  \qsq interval  [\gevgevcccc] & &\multicolumn{4}{c}{ $\frac{\deriv\BF(\decay{\Lb}{\Lz\mumu})/\deriv\qsq}{\BF(\decay{\Lb}{\jpsi\Lz})} \cdot 10^{-3} [(\gevgevcccc)^{-1}]$} \\
\hline
0.1 -- 2.0   & &0.56 & $^{+0.20}_{-0.17}$ & $^{+0.03}_{-0.03}$ & \\
2.0 -- 4.0   & &0.18 & $^{+0.18}_{-0.15}$ & $^{+0.01}_{-0.01}$ & \\
4.0 -- 6.0   & &0.04 & $^{+0.14}_{-0.04}$ & $^{+0.01}_{-0.01}$ & \\
6.0 -- 8.0   & &0.40 & $^{+0.20}_{-0.17}$ & $^{+0.01}_{-0.02}$ &\\
                                                 
11.0 -- 12.5 & &1.19 & $^{+0.24}_{-0.23}$ & $^{+0.04}_{-0.07}$& \\
15.0 -- 16.0 & &1.78 & $^{+0.31}_{-0.28}$ & $^{+0.08}_{-0.08}$&\\
16.0 -- 18.0 & &1.94 & $^{+0.23}_{-0.22}$ & $^{+0.04}_{-0.09}$&\\
18.0 -- 20.0 & &1.97 & $^{+0.23}_{-0.22}$ & $^{+0.10}_{-0.07}$&\\
              
\hline        
1.1--6.0   & &0.14 & $ ^{+0.10}_{-0.09}$& $^{+0.01}_{-0.01}$&\\
15.0--20.0 & &1.90 & $ ^{+0.14}_{-0.14}$& $^{+0.04}_{-0.06}$&\\
\end{tabular}
\label{tab:RelBR}
\end{table}

\newpage

\begin{figure}[htbp]
\centering
\includegraphics[width=0.49\textwidth]{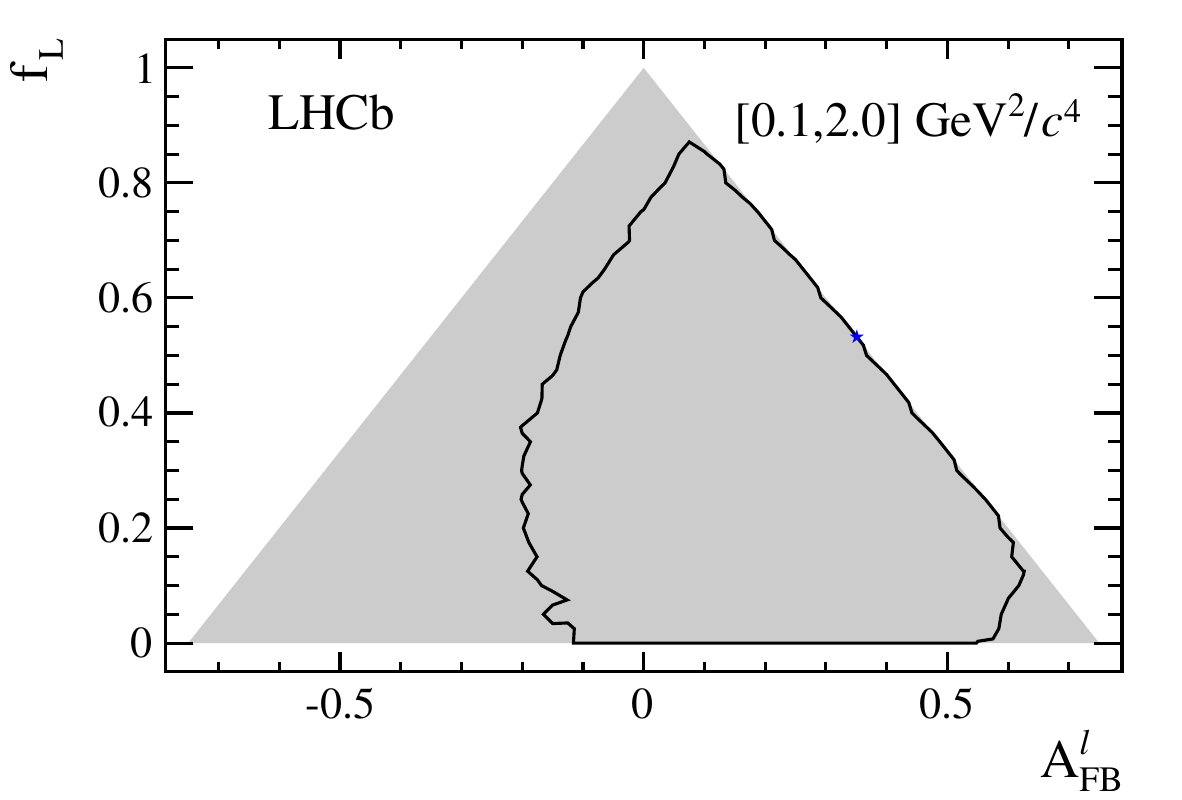}
\includegraphics[width=0.49\textwidth]{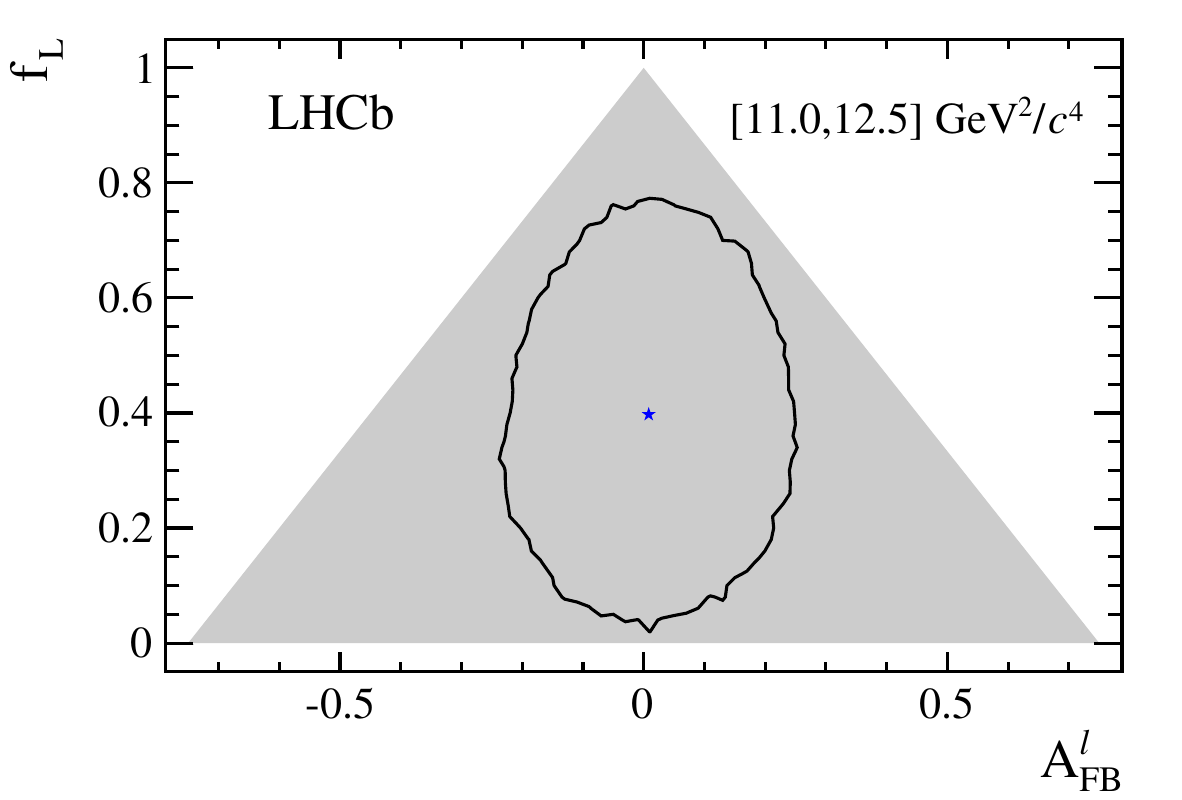}
\includegraphics[width=0.49\textwidth]{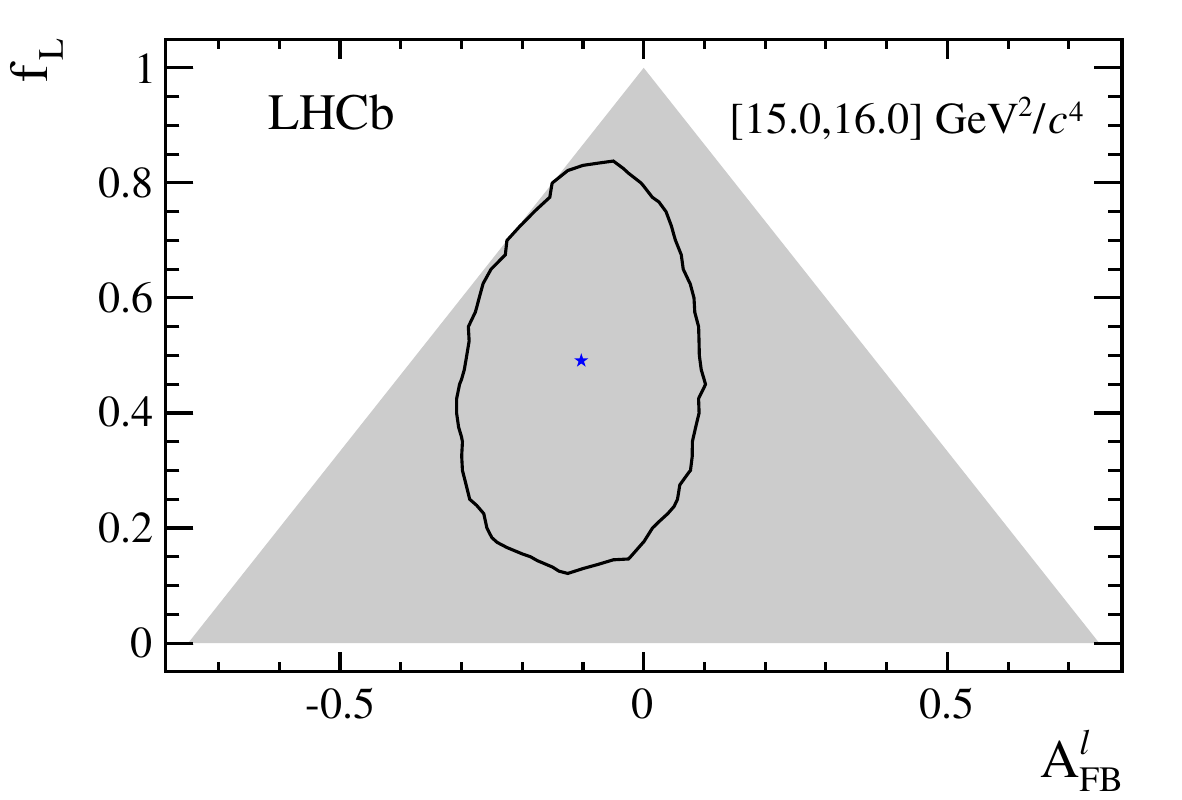}
\includegraphics[width=0.49\textwidth]{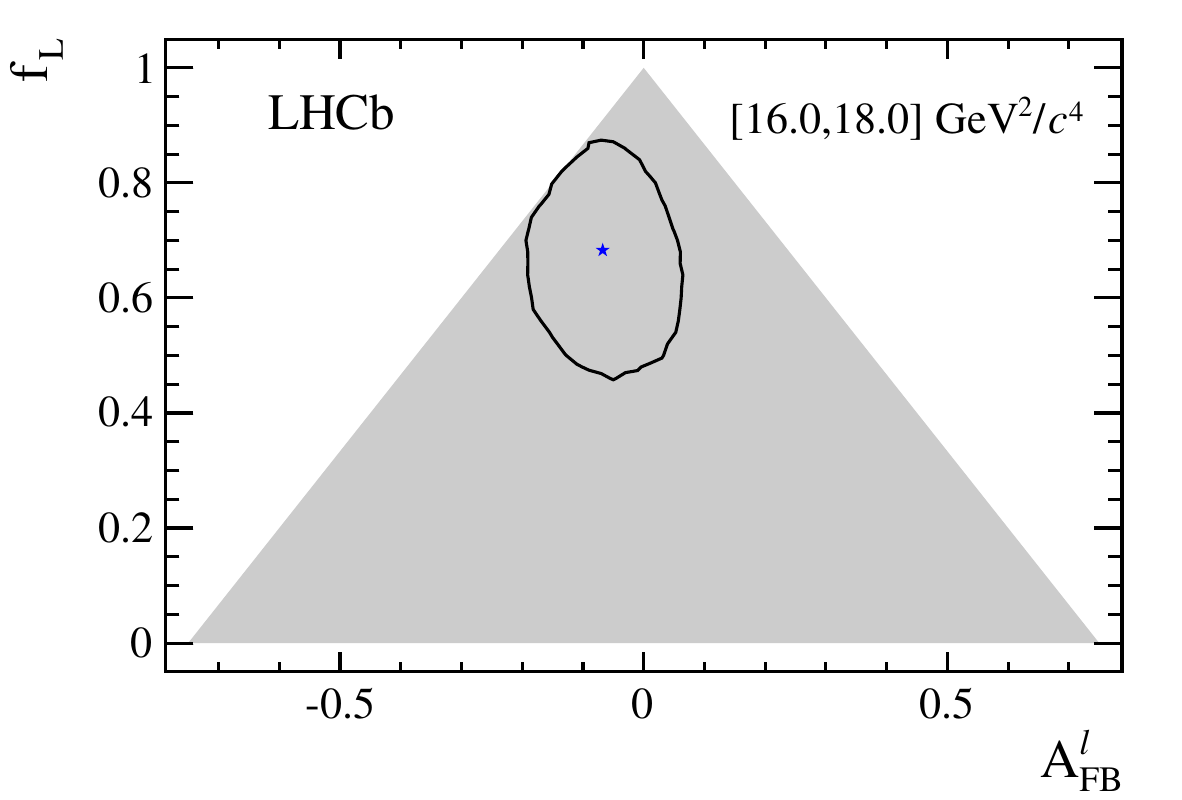}
\includegraphics[width=0.49\textwidth]{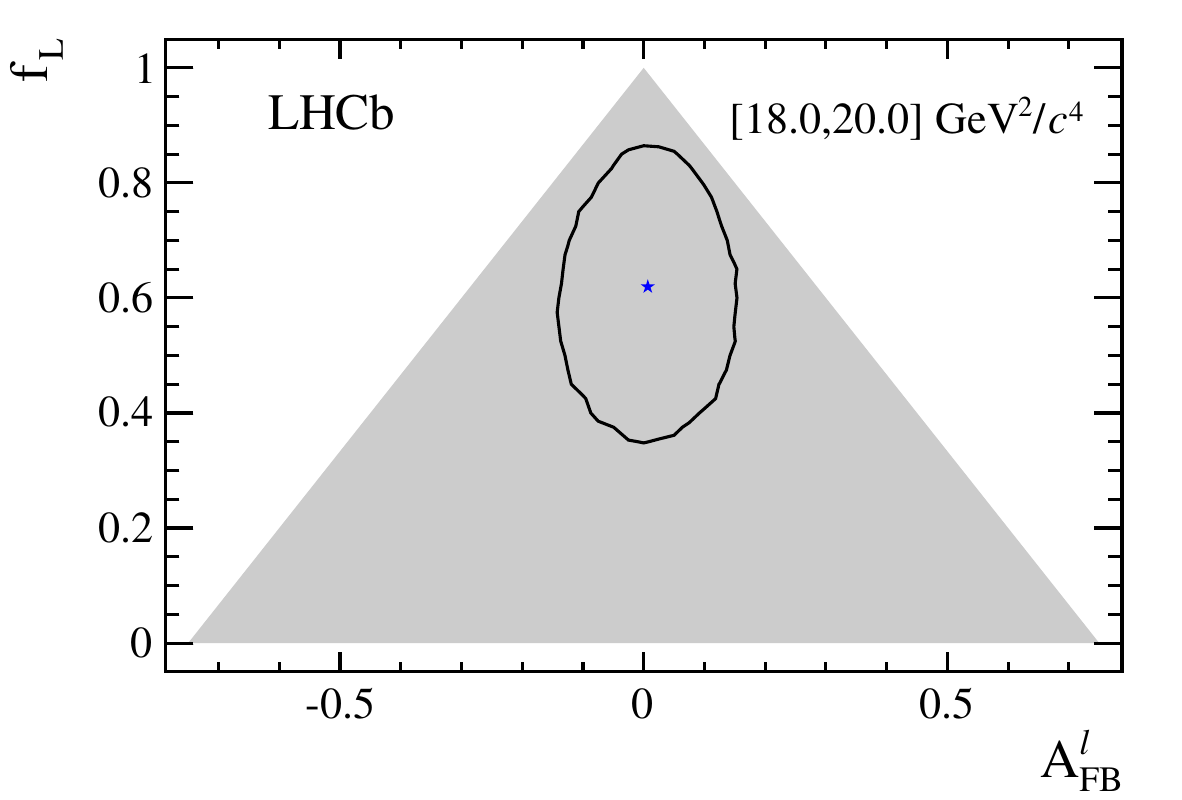}
\caption{Two-dimensional 68\,\% CL regions (black) as a
  function of $A_{\rm FB}^\ell$ and $f_{\rm L}$.  The shaded areas
  represent the regions in which the PDF is positive over the complete $\cos
  \theta_{\ell}$ range. The best fit points are indicated by the (blue) stars. }
\label{fig:contours}
\end{figure}

The two-dimensional 68\,\% CL regions for the observables 
$A_{\rm FB}^\ell$ and $f_{\rm L}$ are given in Fig~\ref{fig:contours},
for each \qsq interval in which signal is observed.

\newpage
\section{Erratum}
\label{sec:erratum}

The angular distribution of the dimuon system of the decays \decay{\Lb}{\Lz\mumu} and \mbox{\decay{\Lbbar}{\Lbar\mumu}} can be described by
\begin{align}
\frac{\deriv\Gamma}{\deriv\cos\thetal} = \frac{3}{8}(1 + \cos^2 \thetal)( 1 - f_{\rm L} ) + \AFBl \cos\thetal  + \frac{3}{4} f_{\rm L}\sin^{2}\thetal~,
\end{align} 
where \AFBl is the forward-backward asymmetry of the dimuon system and $f_{\rm L}$ is its longitudinal polarisation fraction. 
For the \Lb decay, the angle \thetal is calculated as the angle between the direction of the \mup lepton, in the rest frame of the dimuon pair, and the direction of the dimuon pair, in the rest frame of the \Lb decay. 
The forward-backward asymmetry of the lepton pair, \AFBl, is ``odd''
under \CP conjugation and changes in sign between the \Lb and \Lbbar decays. 
To compensate for this sign, the angle \thetal is usually calculated from the \mun lepton rather than the \mup lepton such that \AFBl can be calculated from the combined sample. 
This was the intended approach of this paper. 
Unfortunately, \AFBl was determined using the \mup lepton when determining \thetal for both the \Lb and the \Lbbar decays. 
Consequently, the value of \AFBl in this paper corresponds to a difference {$A(\AFBl)$} in asymmetries between the \Lb and \Lbbar decays rather than a proper average and is expected to be zero if \CP is conserved. 
{The result quoted as $\AFBl$ in this paper should therefore be interpreted as
\begin{align}
A(\AFBl) = -0.05\pm0.09\,({\rm stat}) \pm 0.03\,({\rm syst})~,
\end{align}
and} is indeed consistent with the Standard Model expectation that \CP violating effects should be small in the decay \decay{\Lb}{\Lz\mumu}. 
This is in itself a useful result. 
A  measurement of \AFBl has since been presented in Ref.~\cite{LHCB-PAPER-2018-029}. 
The results in Ref.~\cite{LHCB-PAPER-2018-029} supersede the corresponding results in this paper.
Note, the mistake in the angular definition only affects the value of \AFBl presented in the paper. 
The values of \AFBh and the differential branching fraction are unchanged, due to the symmetry of the efficiency model in $\cos\thetal$.

\newpage


\section*{Acknowledgements}

\noindent We express our gratitude to our colleagues in the CERN
accelerator departments for the excellent performance of the LHC. We
thank the technical and administrative staff at the LHCb
institutes. We acknowledge support from CERN and from the national
agencies: CAPES, CNPq, FAPERJ and FINEP (Brazil); NSFC (China);
CNRS/IN2P3 (France); BMBF, DFG, HGF and MPG (Germany); INFN (Italy); 
FOM and NWO (The Netherlands); MNiSW and NCN (Poland); MEN/IFA (Romania); 
MinES and FANO (Russia); MinECo (Spain); SNSF and SER (Switzerland); 
NASU (Ukraine); STFC (United Kingdom); NSF (USA).
The Tier1 computing centres are supported by IN2P3 (France), KIT and BMBF 
(Germany), INFN (Italy), NWO and SURF (The Netherlands), PIC (Spain), GridPP 
(United Kingdom).
We are indebted to the communities behind the multiple open 
source software packages on which we depend. We are also thankful for the 
computing resources and the access to software R\&D tools provided by Yandex LLC (Russia).
Individual groups or members have received support from 
EPLANET, Marie Sk\l{}odowska-Curie Actions and ERC (European Union), 
Conseil g\'{e}n\'{e}ral de Haute-Savoie, Labex ENIGMASS and OCEVU, 
R\'{e}gion Auvergne (France), RFBR (Russia), XuntaGal and GENCAT (Spain), Royal Society and Royal
Commission for the Exhibition of 1851 (United Kingdom).

\addcontentsline{toc}{section}{References}
\setboolean{inbibliography}{true}
\bibliographystyle{LHCb}
\bibliography{main,LHCb-PAPER,LHCb-CONF,LHCb-DP,LHCb-TDR}

\newpage

\centerline{\large\bf LHCb collaboration}
\begin{flushleft}
\small
R.~Aaij$^{41}$, 
B.~Adeva$^{37}$, 
M.~Adinolfi$^{46}$, 
A.~Affolder$^{52}$, 
Z.~Ajaltouni$^{5}$, 
S.~Akar$^{6}$, 
J.~Albrecht$^{9}$, 
F.~Alessio$^{38}$, 
M.~Alexander$^{51}$, 
S.~Ali$^{41}$
% ...

{\footnotesize \it
% ...
$ ^{5}$Clermont Universit\'{e}, Universit\'{e} Blaise Pascal, CNRS/IN2P3, LPC, Clermont-Ferrand, France\\
$ ^{6}$CPPM, Aix-Marseille Universit\'{e}, CNRS/IN2P3, Marseille, France\\
% ...
$ ^{9}$Fakult\"{a}t Physik, Technische Universit\"{a}t Dortmund, Dortmund, Germany\\
% ...
$ ^{37}$Universidad de Santiago de Compostela, Santiago de Compostela, Spain\\
$ ^{38}$European Organization for Nuclear Research (CERN), Geneva, Switzerland\\
% ...
$ ^{41}$Nikhef National Institute for Subatomic Physics, Amsterdam, The Netherlands\\
% ...
$ ^{46}$H.H. Wills Physics Laboratory, University of Bristol, Bristol, United Kingdom\\
% ...
$ ^{51}$School of Physics and Astronomy, University of Glasgow, Glasgow, United Kingdom\\
$ ^{52}$Oliver Lodge Laboratory, University of Liverpool, Liverpool, United Kingdom\\
% ...
}
\end{flushleft}


\end{document}